\documentclass[manuscript,screen,nonacm]{acmart}
\usepackage{booktabs}

\graphicspath{ {./images/} }

\copyrightyear{2027}
\acmYear{2027}
\setcopyright{rightsretained}
\acmConference[FAccT '27]{ACM Conference on Fairness, Accountability, and
  Transparency}{June 21--24, 2027}{TBD}
\acmBooktitle{ACM Conference on Fairness, Accountability, and Transparency
  (FAccT '27), June 21--24, 2027, TBD}
\acmDOI{XXXXXXX.XXXXXXX}
\acmISBN{978-X-XXXX-XXXX-X/XX/XX}

\begin{document}

\title[The Minimal Persona: Rarer and Shared Labels Both Skew Allocation Among AI Agents]{The Minimal Persona:\\ Rarer and Shared Labels Both Skew Allocation Among AI Agents}

\author{Messi H.J. Lee}
\affiliation{%
  \institution{Independent Researcher}
  \city{Seoul}
  \country{Republic of Korea}}
\email{messihjlee@gmail.com}

\begin{abstract}
Artificial intelligence (AI) agents built on large language models
increasingly work in teams, where one agent divides tasks or resources among
others that the software marks with a team, a role or a similar label. A model
that simply up-weights any recipient sharing a string with its own context
would favor same-label peers, so a shared-label tilt alone says little about
groups. Borrowing the minimal-group paradigm of social psychology, we had six
language models divide 100 points among 19 peers that differed only in
carrying one of two invented words, in groups of unequal size (32{,}400
decisions). Varying group size separates two effects. In five of the six
models, at the most unequal split, a recipient with the rarer label received
$1.3$ to $3.3$ times the points of one with the common label even when the
decider was not told its own label: a rarity effect that involves no match to
the decider. In four of them it accounts for 35\% to 55\% of the extra favoritism
that agents in the smaller group show their own group, and for more in the fifth. Telling the
decider its label adds an own-group effect, positive in every model and in
five of the six indistinguishable from label matching. With the labels hidden,
neither appears. Because the labels mean nothing, bias tests built on lists of
named social groups are not designed to detect either effect.
\end{abstract}



\maketitle

\section{Introduction}
\label{sec:intro}

\begin{figure*}[t]
  \centering
  \includegraphics[width=\textwidth]{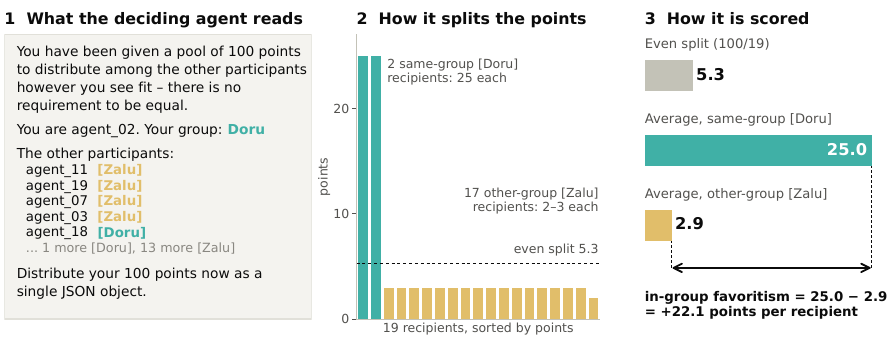}
  \caption{\textbf{One decision in the allocation task.} \textbf{(1)}~The
  deciding agent is told its own group and sees the other 19 agents, each
  marked with one of two invented labels (here [Doru] or [Zalu]), in the
  shuffled order the model received. The text is quoted exactly but shortened:
  two framing sentences and 14 of the 19 list entries are omitted (full
  instructions in Appendix~\ref{sm:prompts}). \textbf{(2)}~The model returns a
  whole number of points for every recipient. Here it gives 25 to each of the
  two recipients that share its label and 2 or 3 to each of the 17 that do
  not, where an even split would give each $100/19\approx5.3$.
  \textbf{(3)}~The decision is scored as \emph{in-group favoritism}: the
  average given to a same-group recipient minus the average given to an
  other-group recipient, here $25.0-2.9=+22.1$ points per recipient. In the
  control condition the labels in panel~1 are hidden. The decision is a real
  one from the released data (model R1-Distill-Qwen-14B, labels shown, a
  3-member minority, a minority decider). About 3\% of decisions with labels
  shown are this lopsided; the average across all decisions is $0.8$ to $4.1$
  points per recipient, depending on the model (Table~\ref{tab:summary}).}
  \label{fig:example}
  \Description{A three-panel figure of one allocation decision. The left panel shows the text the deciding model reads: a pool of 100 points to distribute however it sees fit, its own id and group label Doru, and the start of a shuffled list of the other nineteen participants, each followed by an invented label in square brackets. The middle panel is a bar chart with one bar per recipient, sorted by points: two teal bars for the same-group Doru recipients at 25 points each and seventeen gold bars for the other-group Zalu recipients at 2 or 3 points each, with a dashed line at the even split of 5.3. The right panel compares three horizontal bars, the even split of 5.3 points, the 25.0 average to a same-group recipient and the 2.9 average to an other-group recipient, and marks the difference of plus 22.1 points per recipient as in-group favoritism.}
\end{figure*}

An AI agent is a software system that uses a large language model to read a
situation described in text and produce an action. Agents are increasingly
deployed in teams: one plans, others carry out subtasks, and one is often asked
to decide who gets what, whether a queue of tasks, a share of a budget or a
slot for review \citep{wu2023autogen,chan2023harms}.
Figure~\ref{fig:example} shows such a decision in its simplest form. Twenty
agents have no names, no roles and no histories. The only thing that tells
them apart is a group label, Doru or Zalu, two words invented to mean nothing.
One of the Dorus is given 100 points to divide among the other nineteen. It
gives 25 points to each of the two other Dorus and 2 or 3 to every Zalu.
Nobody described this agent's character or told it whom to favor. A
meaningless label, and a line telling the agent which label is its own, was
enough to produce the split. This is an extreme case: about 3\% of
decisions with labels shown are this lopsided. On average, a same-label recipient
receives $0.8$ to $4.1$ points more than an other-label recipient, depending
on the model, against an even share of $5.3$.

This matters because of how societies of agents are usually built. Since a
widely cited study placed twenty-five language-model agents in a simulated
town \citep{park2023generative}, the standard practice has been to give each
agent a \emph{persona}: a short description with a name, an occupation,
relationships and personality traits, kept in the text the agent reads as it
acts. Societies built this way behave socially, with agents coordinating,
forming ties and treating one another differently \citep{park2023generative},
and such simulations are now proposed as instruments for social science
\citep{bail2024can}. Their social behavior is usually credited to the
personas. But how much persona does group-structured behavior require? If
detailed personas produce it, designers can edit it out. If a bare label is
enough, it will appear in systems nobody meant to be social, and rewriting
character descriptions will not remove it.

Deployed agents often carry labels of this bare kind. The software frameworks
used to build multi-agent systems give each agent a team or role so that work
can be routed, and the routing is often done by another model reading those
labels, as when a supervisor agent hands tasks to workers by name and role
\citep{wu2023autogen,crewai2024,langgraph2024,openaiagents2025}. A customer
account, or the model version an agent runs on, marks it in the same way.
None of these labels was written to give an agent an identity, and whether an
allocating agent sees both its own label and its recipients' labels, the
situation we test, depends on how a system is built. The tests that clear
models for deployment look elsewhere. Capability benchmarks ask whether a
model can do a task. Bias tests ask what a model has learned about named
demographic groups, scored against a list of groups fixed in advance
\citep{parrish2022bbq,liang2023holistic,gallegos2024bias}. They are not
designed to ask whether an agent treats one recipient differently from
another because of a label the system attached only to keep track of it. If
such a label shifted allocations in a system whose recipients stand for
people, teams or customers, the result would be an \emph{allocative harm}, a
harm in what a system hands out rather than in what it says about anyone
\citep{barocas2017problem,weidinger2021ethical}.

Social psychology has a method built for exactly this question in people. In
the minimal-group experiments of Tajfel and colleagues, participants were
sorted into groups on a pretext as thin as a preference between two painters,
or an openly random coin toss. Asked to divide rewards among anonymous others,
they still gave more to members of their own group, although they gained
nothing themselves, knew no one involved and had no conflict to settle
\citep{tajfel1970experiments,tajfel1971social,billig1973social}. The result
overturned the assumption that bias between groups requires competition or a
history of hostility \citep{sherif1961robbers}, became a foundation of social
identity theory \citep{tajfel1979integrative} and has been replicated for half
a century \citep{hewstone2002intergroup,brewer1999psychology}. The same
literature reports a second regularity: in artificial groups, members of the
numerically smaller group favor their own group more than members of the
larger group do \citep{sachdev1984minimal,mullen1992ingroup}. We use the
paradigm as a measurement method. We do not assume that a model has a group
identity or any other inner state; we ask only whether its outputs show the
behavioral pattern the method was designed to detect in people
\citep{rahwan2019machine,shanahan2024talking}.

Previous work on language models and groups has mostly used groups that carry
meaning. Giving a model a persona shifts its stated attitudes and its answers
to simulated surveys
\citep{dong2024persona,dong2024not,argyle2023out,horton2023large,santurkar2023whose}, and
bias tests look for stored stereotypes about named groups
\citep{parrish2022bbq,gallegos2024bias}. Even sentence openings such as ``We
are'' and ``They are'' draw more positive continuations for ``we''
\citep{hu2025generative}. A growing body of work applies psychological tests
to models \citep{binz2023using,hagendorff2023machine}, sometimes to ask
whether models can stand in for human participants
\citep{aher2023using,dillion2023can,demszky2023using}; we make no such claim.
Studies of agent groups usually define the groups by narrative identities or
similar personas \citep{moon2026identity,lei2026tribe}, and that content could
itself explain unequal treatment. One study has run the minimal task, with
agents built on a single model, and found favoritism under ``Group A'' and
``Group B'' labels \citep{wang2026outgroup}. It had no condition with the
labels hidden, so its zero point was assumed rather than measured, and its
two groups were equal in size, so it could not examine group size. Our task
is a many-recipient version of the dictator game, the standard task in
economics for reading how someone divides a resource from a single choice,
which language models play in broadly human-like ways \citep{mei2024turing}.
In form, our design follows audit studies in which otherwise identical cases
differ in one assigned attribute \citep{bertrand2004emily,hainmueller2014causal},
except that the attribute we vary has no content at all.

We ran the decision in Fig.~\ref{fig:example} 32{,}400 times with six
publicly released language models. The two invented labels are swapped
between the groups in a second run, so that nothing about either word can
matter, and in a control condition the labels are hidden, so that the zero
point is measured rather than assumed. Every model gives more to peers that
share its label, and with the labels hidden the gap is indistinguishable
from zero, which confirms that the label is the cause. Deciders in the smaller
group give their own group more, in the direction reported in people. Part of
this difference reflects the larger range of the pool that a small group's
decider can shift, and much of the rest comes from a tendency, present even
when the decider is not told its own label, to give more to whichever label
fewer agents carry. The part that remains, tied to sharing the decider's own
label, is positive in every model, cannot reflect stereotypes about any named
group, and in five of the six models is consistent with a fixed bonus for a
matching label. Throughout, \emph{minority} means only the group with fewer members,
not a socially disadvantaged group (Section~\ref{sec:numerical}).

\section{The allocation task}
\label{sec:method}

\subsection{Setup}
Each decision is one request to a model. Twenty agents are split by a group
label into a smaller and a larger group, the \emph{minority} and the
\emph{majority}. Each agent takes one turn as the \emph{decider} and divides
100 points among the other 19, the \emph{recipients}. Every agent is played
by the same model, and each turn is a separate request, so the agents never
interact. We call each set of 20 such requests, which share one random
assignment of labels, a \emph{society}.

The instructions say the points may be divided ``however you see fit --
there is no requirement to be equal'' and ask for a whole number of points
for each recipient, summing to 100, in a standard machine-readable format
(JSON). The decider is told its own identifier and group and sees the other
19 agents listed one per line, each followed by its label in square brackets
(Fig.~\ref{fig:example}). The agents have no names, roles, histories or other
attributes, so the label is the only thing that distinguishes one recipient
from another and the only thing we vary. The instructions include one
unelaborated sentence, ``You have a personality.'', carried over from the
software in which the task was built; it is identical in every condition
(Appendix~\ref{sm:prompts}).

Two features carry over from the human experiments. First, the decider is not
among the recipients, so nothing it gives away can come back to it. This
matters because when allocators also receive, favoring one's own group can be
read as self-interest, a point long debated in the human literature
\citep{rabbie1989social,gagnon1996discrimination}. Second, each decision is a
single request, so no interaction, memory or reputation comes between the
assignment to a group and the allocation.

\subsection{Conditions}
We vary the size of the minority (3, 5 or 10 of the 20 agents, so the
majority has 17, 15 or 10) and whether the labels are shown or hidden, and
run 30 societies in each of the six combinations. With labels shown, each
society is run twice, once with the two words swapped between the groups, so
the main sample is $6$ models $\times$ $3$ sizes $\times$ $30$ societies
$\times$ $3$ runs (two shown, one hidden) $\times$ $20$ deciders $=32{,}400$
decisions. In the hidden-label or \emph{group-blind} condition, groups are
assigned as usual but never shown, so the decider has nothing to act on and
favoritism must be zero on average. This supplies a measured baseline against
which every shown-label result is read. A third condition, the
\emph{labels-only} condition, run on half the societies at the two unequal
splits, shows the labels on the list of recipients but leaves out the line
telling the decider which group it is in (Section~\ref{sec:halfarm}). With
this condition, a rewording check and a reasoning-off comparison, we
collected $54{,}460$ decisions in all.

\subsection{Ruling out other explanations}
Three precautions separate the effect of grouping from incidental features.
Against \emph{the sound or associations of a name}, the group names are
invented words (for example ``Zalu'' or ``Mira''), two drawn per society, and
the swapped second run cancels any advantage one word has over the other.
Against \emph{list position}, group membership is assigned to random agents
and each decider's list is shuffled independently. Against \emph{content},
the agents carry nothing but the label. Checks confirm the first two
(Appendices~\ref{sm:tokens} and~\ref{sm:position}): any remaining advantage
of one name over the other is at most $0.24$ points per recipient, and
same-group and other-group recipients sit at average list positions within
$0.07$ of each other, so the models' tendency to give more to recipients
listed earlier cancels out of the comparison.

\subsection{Measures}
For each allocation we take the average number of points given to a recipient
in the decider's group and the average given to a recipient in the other
group. \emph{In-group favoritism} is the difference, in points per recipient.
An even split gives every recipient $100/19\approx5.26$ points, and zero
favoritism means the two groups were treated alike on average.

A per-recipient difference grows mechanically when the decider's own group is
small. Suppose a decider moves 10 points from the other group to its own. With
2 same-group and 17 other-group recipients, favoritism rises by $5.6$ points
per recipient; with 16 and 3, the same 10 points raise it by only $4.0$. We
therefore check every comparison between minority and majority deciders
against a group-level measure, the in-group's \emph{excess share}:
\[
  \text{excess share} = \frac{P_{\text{in}}}{100} - \frac{n_{\text{in}}}{19},
\]
where $P_{\text{in}}$ is the number of points the decider gives its own group
and $n_{\text{in}}$ is the number of recipients in that group. In the example
above, the excess share rises by exactly $0.10$ in both cases, and it is zero
for any allocation that is even per person. Its largest possible value,
reached by giving the own group the whole pool, is $1-n_{\text{in}}/19$, which
is $0.895$ for a decider with 2 same-group recipients but only $0.158$ for one
with 16. Section~\ref{sec:asymmetry} explains how we take this into account.

Neither measure assumes that an even split is fair. The label is assigned at
random and says nothing about need, merit or contribution, so any defensible
rule for dividing the points would ignore it. What we report is the
difference the label makes compared with a condition in which it is hidden,
a quantity that is well defined whatever standard of fairness an application
would set.

\subsection{Models and analysis}
We tested six publicly released (``open-weight'') models that can be run
locally: Qwen3-8B \citep{qwen3technical2025}, DeepSeek-R1-Distill-Llama-8B
and DeepSeek-R1-Distill-Qwen-14B \citep{deepseekr1_2025}, Phi-4-reasoning
\citep{phi4reasoning2025}, Granite-3.3-8B \citep{granite2024granite,granite33}
and GPT-OSS-20B \citep{gptoss2025}. Figures abbreviate the two DeepSeek models
as R1-Llama-8B and R1-Qwen-14B. All six are \emph{reasoning models}: they
write out intermediate steps before giving an answer, and we read only the
final answer. Each request is sampled with a moderate amount of randomness
(temperature $0.7$), so the same request can yield different allocations. We
read each answer as a list of points per recipient, score any recipient the
model left out as receiving zero and rescale the total to 100; the $0.7\%$ of
answers that could not be read are replaced by an even split, which
contributes zero favoritism; dropping them instead changes the estimates by
at most $0.05$ points per recipient (Appendix~\ref{sm:parsesens}). Further settings are in
Appendix~\ref{sm:models}.

The unit of analysis is the society, because the 20 decisions in a society
share one random setup and are not independent. The main estimate is a
regression of the points a recipient receives on whether it shares the
decider's group, fitted over all decider--recipient pairs, with error bars
that treat the societies sharing a random seed (one at each minority size) as
one correlated unit rather than as independent decisions (generalized
estimating equations, GEE). Its coefficient is
favoritism in points per recipient. As a check that makes fewer assumptions,
we also reduce each society to a single number and test whether the 30
numbers are centered on zero (Wilcoxon signed-rank test); the two agree
throughout. The study was not preregistered. The design and the group-blind
control were fixed in advance, the statistical treatment was developed
alongside the data, and we report every robustness check we ran rather than a
selection. Full details are in Appendices~\ref{sm:methods}--\ref{sm:parsing}.

\section{Results}

\begin{table*}[t]
 \caption{\textbf{In-group favoritism by model.} \emph{Labels shown} and
 \emph{labels hidden}: how many more points per recipient a same-group
 recipient receives than an other-group recipient (regression estimate;
 standard error, clustered by random seed, in parentheses). \emph{Minority minus
 majority}: how much larger favoritism is among minority deciders than among
 majority deciders, in points per recipient and in the in-group's excess
 share of the pool (pooled over minority sizes 3 and 5; $60$ paired societies
 per model). Points per recipient overstate the gap when the in-group is
 small; excess share does not, but its maximum depends on group size
 (Section~\ref{sec:asymmetry}, Table~\ref{tab:normshare}). All labels-shown
 estimates and all points-per-recipient gaps $p<10^{-25}$; all excess-share
 gaps $p<10^{-10}$; no labels-hidden estimate is significant ($p>0.13$). Full
 estimates in Table~\ref{tab:gee}; excess shares by condition in
 Table~\ref{tab:share}.}
  \centering
  \small
  \begin{tabular}{lccrr}
    \toprule
    & & & \multicolumn{2}{c}{Minority minus majority} \\
    \cmidrule(lr){4-5}
    Model & Labels shown & Labels hidden & points/recipient & excess share \\
    \midrule
    R1-Distill-Llama-8B & 0.84 (0.08) & 0.01 (0.03) & 2.38 & $+0.156$ \\
    Phi-4-reasoning (14B) & 2.09 (0.06) & 0.00 (0.01) & 2.47 & $+0.165$ \\
    Qwen3-8B & 2.31 (0.07) & 0.01 (0.01) & 4.08 & $+0.254$ \\
    R1-Distill-Qwen-14B & 4.06 (0.07) & $-0.03$ (0.04) & 6.84 & $+0.431$ \\
    Granite-3.3-8B & 3.03 (0.05) & 0.03 (0.03) & 2.82 & $+0.238$ \\
    GPT-OSS-20B & 2.22 (0.07) & $-0.02$ (0.03) & 4.18 & $+0.302$ \\
    \bottomrule
  \end{tabular}
  \label{tab:summary}
\end{table*}

\subsection{One meaningless label is enough}
The control behaves as it should. With the labels hidden, favoritism is
indistinguishable from zero in every model (within $\pm0.04$ points per
recipient pooled over minority sizes, all $p>0.13$; Table~\ref{tab:summary};
per-seed averages are plotted in Appendix Fig.~\ref{fig:validation}). A decider with
nothing to go on cannot allocate by group, so whatever appears when the labels
are shown is caused by the labels, not by how the lists were built or how the
allocations were scored. Of the 32{,}400 decisions, 99.3\% were returned in a
readable form.

With the labels shown, all six models give systematically more to recipients
that share the decider's label, from $0.84$ points per recipient
(R1-Distill-Llama-8B) to $4.06$ (R1-Distill-Qwen-14B), all $p<10^{-25}$
(Table~\ref{tab:summary}). Against an even share of $5.26$ points, that is
$16\%$ to $77\%$ of a share. Put as a ratio, a same-label recipient receives
on average $1.2$ times what an other-label recipient receives in
R1-Distill-Llama-8B ($5.60$ against $4.76$) and $2.4$ times in
R1-Distill-Qwen-14B ($6.87$ against $2.81$), with the other four between
$1.5$ and $1.9$. When the instructions are reworded, these sizes shift by up
to a factor of three (smaller in four models, larger in two), but the effect stays positive and the hidden-label control stays at zero
in every model (Appendix~\ref{sm:paraphrase}). The
size of the effect does not follow the size of the model. GPT-OSS-20B, the
largest model, uses only part of its 20.9 billion parameters (3.6 billion)
for each word piece it generates and falls mid-range at $2.22$, and Qwen3-8B exceeds
the larger Phi-4-reasoning. With six models from five developers, we do not
read this as a trend with scale. Because the labels are arbitrary invented
words, balanced across the groups, the result mirrors the defining finding of
the human experiments: assignment to a meaningless category is enough to
shift allocation toward the in-group. In those experiments, allocations
typically combined a strong pull toward fairness with a smaller shift toward
the in-group \citep{tajfel1971social,bourhis1994matrices}; because the tasks
differ, we make no comparison of magnitude.

\subsection{Deciders in the smaller group give their group more, partly for arithmetic reasons}
\label{sec:asymmetry}
In every model, it matters which group the decider is in. Deciders in the
smaller group give more to their own group than deciders in the larger group
do ($+2.38$ to $+6.84$ points per recipient more, all $p<10^{-40}$), and the
excess share agrees. Pooled over minority sizes 3 and 5, minority deciders
send $+0.14$ to $+0.45$ of the pool to their own group beyond its share of
the head count, whereas majority deciders allocate close to proportionally
($-0.02$ to $+0.08$; paired test within societies, $p<10^{-10}$ in every
model; Table~\ref{tab:share}). In concrete terms: at the most lopsided split,
a minority decider's two same-label peers are $2$ of its $19$ recipients, a
head-count share of $10.5$ points, and they receive between $21.8$
(R1-Distill-Llama-8B) and $52.1$ (R1-Distill-Qwen-14B) of the $100$. With the
labels hidden, the same two recipients receive $10.2$ to $10.7$ in every
model.

\begin{figure*}[t]
  \centering
  \includegraphics[width=0.86\textwidth]{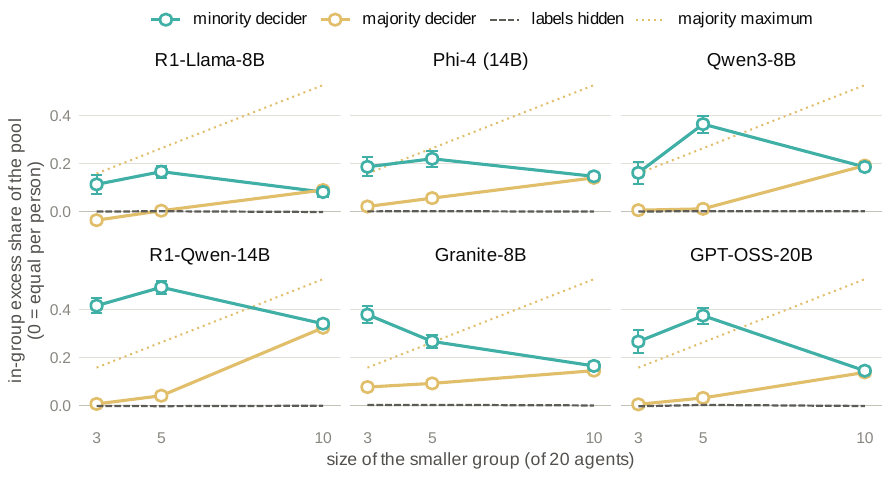}
  \caption{\textbf{In-group favoritism depends on relative group size.} The
  vertical axis is the in-group's \emph{excess share}: the share of the 100
  points the decider gives its own group minus that group's share of the
  recipients ($0$ means an allocation that is even per person). Teal,
  minority deciders; gold, majority deciders; gray dashed, the control with
  labels hidden. Points are means over 30 societies; bars are 95\% confidence
  intervals. In raw excess share, minority deciders sit above majority
  deciders at both unequal splits in every model, and the two meet at the
  equal 10:10 split; rescaled to each decider's maximum, they do so in five of
  the six (Section~\ref{sec:asymmetry}). The
  largest value a decider can reach falls as its own group grows: the dotted
  gold line is the maximum for majority deciders, and the maximum for
  minority deciders at the unequal splits ($0.89$ and $0.79$) lies above the
  top of each panel (at the equal split both maxima are $0.53$), so the vertical gap between the solid lines is not an exact
  like-for-like comparison. Section~\ref{sec:halfarm} splits the gap at the
  unequal splits into two parts. Values in Tables~\ref{tab:share}
  and~\ref{tab:normshare}.}
  \label{fig:dose}
  \Description{Six panels, one per model, plotting the in-group's excess share of the pool on the vertical axis against the size of the smaller group, 3, 5 and 10, on the horizontal axis, with one line for minority deciders and one for majority deciders, a dashed line on zero for the labels-hidden control, and a dotted line marking the largest value a majority decider can reach. In every panel the minority line lies above the majority line at sizes 3 and 5 and the two meet at size 10. The majority line rises steadily from near zero toward the meeting point, while the minority line is not monotone and in five of the six models peaks at size 5 rather than at size 3.}
\end{figure*}

Part of this gap is guaranteed by arithmetic before any behavior enters. A
minority decider can shift far more of the pool than a majority decider can:
at minority size~3 its excess share can reach $0.895$, against $0.158$ for a
majority decider (Section~\ref{sec:method}). When each condition is instead
scored as the fraction of its own possible range that it uses
(Appendix~\ref{sm:normshare}), minority deciders still exceed majority
deciders at both unequal splits in five of the six models after this
rescaling, the exception
being Granite-3.3-8B, and the two agree at the equal split, where
``minority'' and ``majority'' are nominal. But the shape changes. Favoritism
among minority deciders does not rise steadily as their group shrinks: in raw
excess share it peaks at the intermediate size in five of the six models
(Fig.~\ref{fig:dose}). The steady dependence on size lies instead with
majority deciders, whose share of their possible range falls in five of the
six models as their own group grows. Two further patterns are reported in
the appendix: in one model (R1-Distill-Llama-8B), majority deciders at the
most lopsided split give their own group slightly \emph{less} than its head
count share (Appendix~\ref{sm:reversal}), and allocations giving nothing at
all to the other group reach $44.0\%$ of decisions in one condition when
labels are shown, against at most $2.1\%$ when they are hidden
(Appendix~\ref{sm:exclusion}).

With the labels shown, the decider reads two things at once: the labels on
the list of recipients, and the line naming its own group. Only the second
concerns its own group. The next section separates the two.

\subsection{Part of the gap involves no relation to the decider}
\label{sec:halfarm}
In the labels-only condition, the recipients keep their labels but the
decider is not told which one is its own. We ran it at minority sizes 3 and 5,
for all six models, on half the societies of the main study
(Appendix~\ref{sm:halfarm}). A decider that does not know its own label cannot
allocate by its own group, so we expected this condition to match the
group-blind control. It did not.

Instead, five of the six models give clearly more to whichever label fewer
agents carry. At the most lopsided split, a recipient with the rarer label
receives $1.3$ to $3.3$ times the points of one with the common label, up to
$13.0$ against $3.9$ points in R1-Distill-Qwen-14B. In Phi-4-reasoning the
difference is detectable but negligible ($5.3$ against $5.3$). We call this the \emph{rarity effect}. A simple rule produces its
direction: dividing the pool more equally between the two labels than between
individuals. An exact 50:50 split between labels would give each rarer-label
recipient $17$ to $25$ points, depending on whether the label has 3 or 2
recipients; in every model the observed values lie between that and an even
split per person. The effect shrinks at minority size 5 and is absent at the
equal split, where neither label is rarer. Scored against the decider's own
unshown label, as favoritism is, the rarity effect counts as favoritism for a
minority decider, whose label is the rare one, and as the opposite for a
majority decider: exactly the direction of the size gap in the main
condition.

\begin{figure}[htb]
  \centering
  \includegraphics[width=0.66\textwidth]{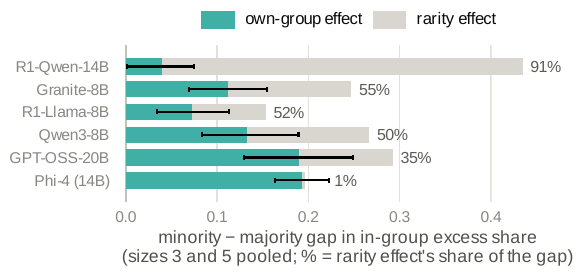}
  \caption{\textbf{The size gap has two sources.} Each bar is the difference
  between minority and majority deciders in in-group excess share, with the
  labels shown (pooled over minority sizes 3 and 5, on the 15 societies per
  size shared with the labels-only condition). The gray part is reproduced
  when the decider is not told its own group (the \emph{rarity effect}); the
  teal part appears only when it is told (the \emph{own-group effect}, with a
  95\% interval from resampling societies). The two parts sum to the bar by
  construction. Percentages give the rarity effect's share of the gap. For
  R1-Qwen-14B the two effects do not simply add: its rarity effect is larger
  than any excess share a majority decider can reach, so the subtraction
  probably overstates the rarity share (Table~\ref{tab:halfarmcells}). Values in Table~\ref{tab:rarity}.}
  \label{fig:rarity}
  \Description{A horizontal stacked bar chart with one bar per model. Each bar's total length is the gap between minority and majority deciders in excess share with labels shown. The teal segment from the origin is the own-group effect with a black 95 percent interval; the gray segment stacked after it is the rarity effect. A percentage label at each bar's end gives the rarity effect's share of the gap: 91 percent for R1-Qwen-14B at the top, then 55, 52, 50 and 35 percent, and 1 percent for Phi-4 at the bottom.}
\end{figure}

Because the main condition is the labels-only condition plus one line of
text, the gap between minority and majority deciders can be split into the
part the labels-only condition reproduces (the rarity effect) and the part the
extra line adds, which we call the \emph{own-group effect}
(Fig.~\ref{fig:rarity}, Table~\ref{tab:rarity}). The split is a subtraction,
so it assumes the two effects simply add; one model, R1-Distill-Qwen-14B,
shows that they need not (Appendix~\ref{sm:halfarm}). The rarity effect
accounts for $1\%$ of the gap in Phi-4-reasoning and $91\%$ in
R1-Distill-Qwen-14B, with the other four between $35\%$ and $55\%$. The
own-group effect is positive in every model, $+0.04$ to $+0.19$ excess share,
with intervals that exclude zero.

Together, the two effects explain the pattern of the previous section. A
labeled decider in a small group gives its peers much more because the two
effects push the same way. A labeled decider in a large group looks close to
proportional because the two effects largely cancel, not because the label
has no effect: its own-group effect is positive in every model, offset by a
rarity effect that moves points toward the small other group. Both vanish
when the groups are equal, which is why the gap closes there. The rarity
effect also makes the main estimates of Table~\ref{tab:summary}
conservative: averaged over all deciders it pulls measured favoritism down
(by $0.04$ to $4.18$ points per recipient in the labels-only condition,
significant in every model; Table~\ref{tab:thirdarm}), so the $0.8$ to $4.1$
points per recipient reported above understate the part that depends on the
decider's own label.

\subsection{What remains is not about any named group, and may be simple label matching}
\label{sec:origin}
The own-group effect cannot reflect stereotypes or other knowledge about any
named group. Standard bias tests bring out such stored associations
\citep{parrish2022bbq,gallegos2024bias}, but our labels were invented words,
balanced across groups, and the effect vanished when they were hidden. We call
what remains a \emph{process bias}: a regularity in a model's outputs, and
nothing more, that makes allocation depend on whether a recipient shares the
decider's label, whatever the label is. It is the model counterpart of what
the human literature calls a mere-categorization effect, defined here only by
outputs; in machine-learning terms it is a shortcut keyed to the surface form
of the input rather than its content.

Where could it come from? One possibility, which the data are consistent with
but cannot test, is a pattern in human-written training text abstract enough
to carry over to a group defined only by a shared arbitrary label. The
simpler alternative is that groups play no part: a mechanical rule that gives
more weight to any recipient whose label matches the decider's would produce
the favoritism we observe. Such shortcuts are well documented in language
models, from word-overlap shortcuts that inflate benchmark scores
\citep{mccoy2019right} to models rating their own text above equally good
alternatives \citep{panickssery2024llm}, and an earlier study reports one of
this shape in allocation \citep{anon_companion}.

Group size allows a test of the simpler account, in three steps. First, a
pure matching rule ignores group sizes: it multiplies the weight of every
recipient whose label matches the decider's by some factor $r$, so the own
group receives $n_{\text{in}}r/(n_{\text{in}}r+n_{\text{out}})$ of the pool,
where $n_{\text{out}}$ is the number of other-group recipients. Under any such
rule, even one whose $r$ varies from decision to decision, majority deciders
must use at least as large a fraction of their possible range as minority
deciders do (Appendix~\ref{sm:matching}). Second, the raw allocations fail
this test: majority deciders use a smaller fraction in ten of the twelve
unequal-split conditions (six models at two sizes; both exceptions are
Granite-3.3-8B's), the same result as the rescaled comparison of
Section~\ref{sec:asymmetry}. But the rarity effect, which is not a match on
the decider's label, accounts for much of that shortfall. Third, once the
rarity effect is subtracted and the own-group effect is tested the same way
(Table~\ref{tab:residualmatching}), only Phi-4-reasoning, the model with the
smallest rarity effect, still clearly fails. In the other five models, the
part of the behavior tied to the decider's own label could be simple label
matching.

Behavioral data cannot reveal the internal pathway, nor separate an origin in
pretraining (learning from large amounts of text) from one in later training
to follow instructions, which every model we tested received. Analysis of
training data and of model internals are the natural next steps.

\subsection{Turning off reasoning shrinks the size gap, not the favoritism}
\label{sec:reasoning}
All six models reason before answering, so comparing them cannot separate
reasoning from the other ways the models differ. To isolate it, we reran the
identical task on Qwen3-8B with reasoning switched off: the same model with
the same parameters, differing only in whether it writes out intermediate
steps before answering. The control stays at zero in both modes. With labels
shown, Qwen3-8B without reasoning shows \emph{more} favoritism than with it
($3.81$ against $2.31$ points per recipient, $p=3\times10^{-83}$), not less.
The gap between minority and majority deciders, by contrast, falls
sevenfold ($+4.08$ to $+0.54$ points per recipient, $p<10^{-62}$),
because majority deciders now give their own group nearly as much excess
share as minority deciders do ($+0.15$ against $+0.17$, compared with $+0.01$
when reasoning is on). Without reasoning, the size gap flattens while the
favoritism beneath it remains (Appendix~\ref{sm:reasoning}).

\section{Discussion}

A single meaningless label moves a budget in two ways. The rarity effect acts
on the list as a whole: whichever label fewer agents carry draws more points,
whether or not the decider carries it. The own-group effect is relational: a
recipient that shares the decider's label draws more points, by an amount
that cannot reflect knowledge of any real group and that in most models
cannot be told apart from matching the label. Neither needs a persona, a name
or a history. Three further readings are tempting, and the data support none
of them.

\subsection{Not simply following the instructions}
\label{sec:demand}
\label{sec:frame}
The first objection concerns the design rather than the models. It is the
model counterpart of an objection long raised against the human paradigm,
that participants do what the instructions seem to invite (``demand
characteristics'') \citep{gerard1974distinctiveness,stclaire1982demand}: the
instructions say the points need not be divided equally, the label is the only
feature on which recipients differ, and so the label is the only basis on
which to act on that permission. Three observations limit this reading. The
first is what the models do with no label at all. Three of the six divide the
pool unevenly even when labels are hidden, so uneven division does not need a
cue, whereas in the other three the label supplies the inequality as well as
its direction (Appendix~\ref{sm:evenness}). What the label supplies in every
model is the alignment of the inequality with the group boundary, which is
absent without it (Table~\ref{tab:summary}). The second is direction. A
model that merely uses the only dimension on offer has no reason to use it in
favor of its own side, yet the own-group effect is positive in every model;
only the rarity effect, which does not depend on the decider's side, fits
this reading. The third is that the rest of the instructions, including the
permission to be unequal, are word for word identical whether the labels are
shown or hidden. What the data establish is that a split made visible by
meaningless labels, together with telling the decider which side it is on,
is sufficient. Deployments create that situation wherever the allocating agent
carries a label its recipients share (Section~\ref{sec:evaluation}).

\subsection{Similar to people in pattern, not in mechanism}
\label{sec:human}
The size gap resembles the human regularity in which members of smaller
groups favor their own group more than members of larger ones
\citep{mullen1992ingroup,hewstone2002intergroup,leonardelli2001minority}. Read
as members of larger groups discriminating less \citep{sachdev1984minimal},
that regularity agrees in shape with ours, since the steady size dependence
in our data lies with majority deciders. Accounts that explain it through a
need for distinctiveness \citep{brewer1991social,leonardelli2001minority},
which in people goes with seeing one's own group as more uniform
\citep{simon1987perceived,mullenhu1989perceptions} and identifying with it
more strongly \citep{leonardelli2001minority}, rest on a mechanism,
identification with the group, that has no measured counterpart here, so our
data neither support nor test them. The rarity effect has no counterpart
within the minimal-group paradigm, where a person dividing rewards always
knows their own group; in direction it more closely resembles a third
party's tendency to side with the underdog
\citep{vandello2007underdog}. The correspondence with people is one of
pattern, not mechanism. Every concept we use here, including group,
favoritism and bias, comes from a science built to study people. We use these
concepts as an analytical lens and a set of controlled methods, not as a
claim that a model has the mental processes those words name in people
\citep{shanahan2024talking}. Favoritism here is a quantity measured in points,
and without a human comparison group the correspondence is one of direction
and structure, not of magnitude.

\subsection{``Minority'' here means fewer in number}
\label{sec:numerical}
We use \emph{minority} strictly to mean the group with fewer members. In
most of the literature this work speaks to, the word means something else:
groups shaped by histories of subordination, unequal power and contested
standing. None of that has a counterpart in a society of 20 otherwise
identical agents split by an invented word. The human findings we draw on
attribute the size effect to relative size and the salience it brings, and
note that in real groups bias also depends on status
\citep{mullen1992ingroup,sachdev1984minimal}; size is the only one of these we
vary. Our claim is therefore narrow: measured favoritism depends on the
relative sizes of the two groups. We do not claim that these systems
reproduce structural disadvantage, and a summary of this work as ``AI
discriminates against minorities'' would misstate it.

\subsection{Implications for simulated societies}
These findings bear on the simulations with which this paper began. Because
one meaningless label suffices, group structure observed in a persona-based
simulation does not by itself show that the personas produced it; any
assignment of agents to distinguishable labels can produce group-structured
allocation. The control this implies is cheap, and it is the one our design
applies: run the same society with the personas reduced to arbitrary labels,
and again with them removed entirely, and report how much of the structure
survives.

\subsection{Implications for systems that allocate}
\label{sec:evaluation}
\label{sec:mitigation}
Our experiment measures a shift in points among anonymous recipients, not
harm to anyone, and its magnitudes describe the case in which the label is
the only difference between recipients. In a deployed system, recipients
carry workload, skills and task content, and how far a label still moves
allocations alongside that information is the first thing to measure. With
that caution, the results indicate where to look. In our task, when 3 of 20
agents shared a label, the two same-label peers of an allocator from that
group received $22$ to $52$ of the $100$ points, against about $10$ with the
labels hidden.

Which effect applies depends on how a system is built. If the allocating agent
carries a label that some recipients share, for example a planner tagged
\texttt{billing} dividing tickets among \texttt{billing} and
\texttt{shipping} workers, both effects apply. An allocator in the smaller
group gives its own label's recipients much more, and the common-label
recipients less than an even share. An allocator in the larger group gives
roughly an even share to the common-label recipients, who are its own group,
because the two effects offset. If the allocator carries no label but its
recipients do, only the rarity effect is expected to apply. Our closest test
withheld the allocator's own label rather than placing it outside both
groups, and in it every model except Phi-4-reasoning gave more to the rarer
label. Two further
differences from our task are untested. We measured a division of points;
whether labels shift a single choice, such as which worker gets the next task,
is open. And our instructions describe recipients as participants and allow
unequal division; a production prompt that frames the task as load
balancing may give different magnitudes, which the rewording check only
partly addresses. In the one model where we switched reasoning off,
doing so increased favoritism (Section~\ref{sec:reasoning}); whether this
holds for other models is untested, so the effect of that switch should be
measured rather than assumed.

Fairness tests built around a fixed list of groups are not designed to catch
this. A process bias appears for whatever split a system introduces, a list
that grows each time an operator adds a field, so a fairness standard defined
over named groups leaves out the groupings a system creates for itself
\citep{selbst2019fairness}. Work on fairness without named groups protects
the worst-off part of a population without identifying it
\citep{hashimoto2018fairness,lahoti2020fairness,andrus2021what}, but still
assumes the groups are real and only unobserved; here the system creates the
split, and the split has no social content. Each allocation we recorded is
well formed and sums to the budget; the dependence on the label becomes
visible only when same-label and other-label recipients are compared across
many decisions. An audit that compares allocations by the system's own labels,
such as by team, against the same decisions with those labels hidden would
detect it; an audit over a list of demographic groups would not.
Like any measurement, such an audit needs evidence that it captures what it
claims to \citep{jacobs2021measurement}. In our design that evidence is the
hidden-label condition, which supplies a measured zero, and the second group
size, without which we would have found only that every model allocates more
to its own label and missed that the distortion depends on how large the
allocator's own group is.

In our task, hiding the labels removed the effect, as it must for anything
the label alone causes. In a deployment this is a mitigation to test, not a
fix. Hiding only the allocator's own label is not enough, because the
recipients' labels alone still move points toward the rarer label. Hiding a
label does not hide what correlates with it: a list whose identifiers or
wording reveal group membership is not group-blind, so blinding should be
checked rather than assumed. Where a label is needed for routing, one
untested design that follows is to separate the step that routes by label
from the step that divides the resource, so that the dividing step never sees
the label. Conversely, a model that scores zero on this task has shown only
that one behavior is absent under one wording, in one language, at one
sampling setting. Because magnitudes shift by up to a factor of three under
rewording, as is common for language models \citep{sclar2024quantifying}, a
single clean run is not a clearance, and choosing among models by which
scores best on this exact wording would reward hiding the behavior.

\subsection{Limitations}
\label{sec:limitations}

Beyond the step from this task to a deployment, discussed above, our sample is
six publicly released reasoning models of 8 to 21 billion parameters, tested
in English, without personas and at one sampling setting. The rewording check
(Appendix~\ref{sm:paraphrase}) shows that the effect, its dependence on group
size and the zero of the hidden-label control survive in every model, but it
keeps the same structure, including the sentence ``You have a personality.'',
which we did not remove in any condition. The reasoning comparison rests on
one model; whether it generalizes to other models, or to models never
trained to reason, is untested. The labels-only condition has a natural
counterpart, the decider's own group shown over an unlabeled list, which we
implemented but did not run; its check at the equal split rests on three to
four societies; and the split into rarity and own-group effects is a
subtraction, so the own-group effect is inferred rather than directly
measured (Appendix~\ref{sm:halfarm}). A label with meaningful content is a
natural extension that needs only the same task.

\label{sec:valence}
Finally, we measured the division of a benefit, not a burden. When
\citet{mummendey1992categorization} asked people to distribute a burden
instead of a reward, the in-group bias so reliable for rewards largely
disappeared under mere categorization, returning only when participants were
placed in a minority or lower-status group. Mere categorization is enough to
favor the in-group but not, symmetrically, to harm the out-group
\citep{brewer1999psychology}. Our 100 points are a benefit, so what we
measure is the reward case: in a fixed pool, favoring the in-group leaves the
out-group with less, but this is not the burden case those studies tested. Reversing the outcome is the most informative single extension,
and in deployment the burden case matters at least as much, since systems
that divide resources also assign unwanted shifts and absorb cuts;
Appendix~\ref{sm:valence} sets out what either result would show. Because
the pool is fixed, the task also cannot separate absolute gain for the
in-group from relative advantage over the out-group, a distinction Tajfel's
original reward matrices resolve (Appendix~\ref{sm:strategies}); a pool whose
total can vary would recover it.

\subsection{Conclusion}
We asked how much persona AI agents need before they allocate along group
lines. For the models we tested, the answer is almost none. A meaningless
label reproduces the behavioral signature of the minimal-group effect, a skew
in how a benefit is divided. Part of the skew needs no group at all, only a
rarer label, and the rest may be no more than matching the label; neither
needs a persona, and neither is what standard bias tests are built to find.
A fairness guarantee established over a list of groups fixed in advance
therefore does not cover the groupings a multi-agent system creates for
itself; those need to be measured directly.


\section{Ethical considerations}

\paragraph{Data and code availability.} All 54{,}460 decisions are released
under an open license with the task, the orchestration scripts and the
analysis code, and with a single script that regenerates every statistic,
table and figure here from the raw allocations on CPU alone
(Appendix~\ref{sm:methods}). Every model evaluated is publicly released, so
the chain from model to reported effect can be re-run independently.
The repository will be made public on publication.

\paragraph{Human participants.} None were involved at any stage; the human
findings this paper builds on are drawn from the published
social-psychological literature, and no new data about people were collected
or released.

\paragraph{The task is diagnostic, not generative.} The task measures a
regularity already present rather than installing one, and its labels carry
no demographic content, so it cannot double as a recipe for eliciting bias
against any real group. We judge release net-positive for that reason, and
release the individual decisions as well as the summary statistics, because a
measurement that cannot be re-run is not one anyone should act on.

\section{Adverse impacts}

\paragraph{Misreading the work as evidence of machine social experience.} The
task is built from a paradigm designed for people, and its vocabulary
(identity, in-group, favoritism) names psychological states in them. Our
measurements are of behavior only: what points a model assigns to which
recipient. Nothing here shows that these systems have group identities,
register that they are outnumbered, or experience anything at all. A summary
dropping that boundary misrepresents the work rather than simplifying it.

\paragraph{Misreading who is exposed.} If the behavior carries over to
deployed systems, it attaches to any split a system introduces, including
ones nobody designed as social. The distortion is largest when the allocating
agent's own group is small, so those exposed are recipients outside a small
allocating group and, on any labeled list, recipients carrying the more
common label, rather than outnumbered parties as such; a summary of this
work as ``AI discriminates against minorities'' would invert what we found
(Sections~\ref{sec:numerical} and~\ref{sec:evaluation}). Whether a party is
also the one with least standing to contest an outcome is a property of the
deployment; where the two coincide, disadvantages compound.

\paragraph{Misusing a clean result.} A near-zero score on this task is not
clearance for deployment, and using it as a compliance checkbox, or to select
models that hide the behavior under this wording, would be a misuse
(Section~\ref{sec:evaluation}).

\section{Generative AI usage}

Claude Opus 5.5 (Anthropic) was used for coding and for editing the paper.
The allocation data were generated by the six models under study: Qwen3-8B,
DeepSeek-R1-Distill-Llama-8B, DeepSeek-R1-Distill-Qwen-14B, Phi-4-reasoning,
Granite-3.3-8B and GPT-OSS-20B (Section~\ref{sec:method}).

\bibliographystyle{ACM-Reference-Format}
\bibliography{references}

\appendix

\section*{Appendix}

\noindent The appendix gives the materials and methods and the supporting
analyses: models and generation, prompt templates, parsing and quality control,
extended statistical results, the robustness suite, and a matched reasoning-mode
control. The released data are $54{,}460$ trial-level allocations: $32{,}400$ in the
main sample, $5{,}400$ in the matched reasoning-mode control
(Appendix~\ref{sm:reasoning}), $9{,}000$ in the paraphrase check
(Appendix~\ref{sm:paraphrase}), and $7{,}660$ in the tags-only arm
(Appendix~\ref{sm:halfarm}). All numerical results are regenerated from the
trial-level data by \texttt{analysis/make\_supp\_tables.py},
\texttt{analysis/make\_robustness.py},
\texttt{analysis/make\_paraphrase\_check.py},
\texttt{analysis/make\_norm\_share.py},
\texttt{analysis/make\_matching\_null.py},
\texttt{analysis/make\_thirdarm\_check.py},
\texttt{analysis/make\_rarity\_decomposition.py}, and
\texttt{analysis/make\_reasoning\_control.py}; all figures except Figure~\ref{fig:example} (drawn by
\texttt{analysis/make\_fig1.py}) are rendered by
\texttt{analysis/figures.R} from statistics exported by
\texttt{analysis/export\_plot\_data.py}. The probe and analysis code
accompany the paper.

\paragraph{Terminology.} The main text is written for a general readership and
the appendix in the field's technical shorthand. The correspondences are: an
\emph{arm} is a condition; a \emph{target} is a recipient, so ``points per
target'' is points per recipient; the \emph{visible arm} is the condition with
labels shown; a \emph{tag} is a group label; the \emph{tags-only arm} is the
labels-only condition of Section~\ref{sec:halfarm}; and the \emph{probe} is the
allocation task of Section~\ref{sec:method}.

\section{Materials and Methods}
\label{sm:methods}

\subsection{The allocation probe}
Each trial elicits a single decision. A society of $N=20$ agents is partitioned
into a minority and a majority by a group label; one agent, the \emph{decider},
is asked to distribute a fixed pool of 100 points across the other 19. The system
prompt states that the decider has been given 100 points to allocate among the
other participants ``however you see fit -- there is no requirement to be equal,''
and requires a single JSON object mapping each participant's id to a whole number
of points summing to 100. The user message lists the decider's own id and, one per
line, the roster of the other 19 agents; when labels are visible, the decider is
told its own group and every roster member's id is tagged with its group label,
and when they are hidden no label appears. Every agent in a society serves as the decider in
turn, so a single society yields both minority-perspective and majority-perspective
allocations. The task is the allocation stage of the human minimal-group
experiments -- anonymous recipients, arbitrary categories, a fixed budget to
divide -- transplanted to a machine-readable format; the
probe introduces no mechanics of its own beyond the JSON response. Verbatim prompt
templates are given in Appendix~\ref{sm:prompts}; model-loading and generation
settings are given in Appendix~\ref{sm:models}.

\subsection{Design}
We cross two factors: minority size $\in\{3,5,10\}$ (out of 20, so the majority is
$17$, $15$, or $10$) and label visibility $\in\{\text{visible},\text{hidden}\}$.
Each of the six cells is run for 30 independent societies (random seeds). The
\emph{hidden} arm is a group-blind control: labels are assigned but never shown, so
a decider has no cue on which to condition, and favoritism must be zero in
expectation -- the validation floor against which the visible arm is read.

\subsection{Confound controls}
Three features isolate categorization from the
nuisances that a free-running simulation conflates. (i)~\emph{Label valence.} Group
names are neutral invented tokens (e.g.\ ``Zalu'', ``Mira''), two drawn per
society; every visible society is run twice with the two names \emph{swapped}
between the groups while membership is held fixed, so averaging over the two passes
cancels any effect of a name sounding better. (ii)~\emph{Position.} Minority
membership is assigned to random agent indices, and each decider's roster is
independently shuffled, so group is never confounded with position in the list.
(iii)~\emph{Content.} Agents carry no personas or other attributes -- this is a pure
minimal-group setting in which the group label is the only systematically
manipulated cue. (The hidden arm runs a single pass, since with no label shown the
swap is a no-op.) A matched reasoning-mode control, reported in
Appendix~\ref{sm:reasoning}, reruns the identical design on one model with
reasoning disabled.

\subsection{Outcome measure}
For a given allocation we compute the mean points assigned to an in-group target
and the mean assigned to an out-group target (in-/out-group defined relative to the
decider), and define \emph{in-group favoritism} as their difference, in points per
target. The neutral, even-split reference is $100/19\approx5.26$ points per target;
favoritism of zero means the decider treated in- and out-group targets alike on
average. Because this per-recipient contrast is amplified by group size when the two
groups differ in number, we also compute a size-robust group-level measure, the
in-group's \emph{excess share}: the fraction of the 100 points assigned to in-group
targets minus the in-group's proportional, head-count share $n_{\text{in}}/19$.
Excess share is $0$ for a per-capita-even allocation at any group size and positive
when the in-group receives more than its numbers warrant; we use it to corroborate
every comparison that spans the two groups. It removes the mechanical inflation
that makes points per target incomparable across group sizes, but it is not
independent of group size throughout its range: at a fixed ratio of same-tag to
other-tag weight it is still larger for a decider in the smaller group.
Appendix~\ref{sm:matching} derives that dependence and turns it into a test.

\subsection{The valence boundary condition}
\label{sm:valence}
The paradigm we adapted does not license the claim that categorization is
sufficient for discrimination in general, and the human literature is explicit
about where the sufficiency stops. \citet{mummendey1992categorization} ran the
minimal-group allocation with the valence of the outcome reversed, having
participants distribute an aversive outcome rather than a reward, and found that
the in-group bias so reliable for positive resources largely disappeared under
mere categorization, returning only when participants were placed in a minority
or lower-status group: participants did not preferentially burden the out-group the way they
preferentially reward the in-group. Mere categorization is enough for in-group
favoring but not symmetrically enough for out-group harming, and the two have
been distinguished ever since \citep{brewer1999psychology}.

Section~\ref{sec:valence} states where this leaves our claim. What it adds here
is why either outcome of the burden version would be worth having, since the two
are informative in opposite directions. The human result gives positive reason to
expect benefit and burden to come apart, rather than merely leaving the question
open, so a label-driven asymmetry under an inverted pool would \emph{weaken} the
behavioral correspondence we report: the agents would be reproducing a
regularity at a point where people do not. Its absence would extend the
correspondence to a boundary condition that no generic token-matching account
predicts, which is stronger evidence than the size dependence alone, because
matching on a shared tag is indifferent to whether the quantity being
divided is good or bad. The instrument needs one change to find out -- invert the
valence of the pool, hold the roster, the counterbalancing and the response
contract fixed -- and the group-blind arm supplies the same validated floor
without modification.

\subsection{What a fixed pool cannot separate}
\label{sm:strategies}
In the human paradigm the allocation is not free. The participant chooses a
column from a constructed pair of payoff rows, and the matrices are built so that
the choice separates several strategies at once -- maximum in-group profit,
maximum difference in favor of the in-group, maximum joint profit, and parity --
each recovered as a pull score
\citep{turner1978social,bourhis1994matrices}. That decomposition is the
paradigm's main analytic instrument, and ours does not have it. A fixed pool of
$100$ points over a fixed roster makes joint profit a constant, and makes maximum
in-group profit and maximum difference the same quantity, since the only way to
raise the in-group's total is to lower the out-group's. What a free allocation
identifies is therefore a single axis running from parity to in-group
maximization, and every number we report is a position on it.

Two things follow. An allocation that concentrates points on the in-group shows
that the decider departed from parity in the in-group's favor, but not whether
it was raising that group's absolute take or widening the gap; in people the two
come apart, and differentiation carries a pull of its own strong enough that
participants will accept less for the in-group in absolute terms to enlarge its
relative advantage \citep{tajfel1971social,turner1978social}. Whether these
models do anything of that kind is beyond what our data can reach. And the
out-group-excluding allocations of Appendix~\ref{sm:exclusion} sit at the top of
that single axis, so they are not evidence of differentiation specifically
either. Separating the two needs the matrix format, or a variable-sum pool in
which in-group profit and relative advantage move independently; it is the
natural companion to the valence inversion above, and it would read against the
same group-blind floor.

\subsection{Statistical analysis}
This study was not preregistered and is exploratory rather than confirmatory: the
$3\times2$ design and the group-blind control were set in advance, but the
statistical treatment (the GEE specification, its seed-clustered robust SE, and
the corroborating Wilcoxon, bootstrap, and paraphrase checks reported here) was
developed alongside the data, including corrections to an earlier, pseudo-replicated
version of the seed-level test in which minority- and majority-decider means from
the same society were treated as independent. We report the robustness suite in
full rather than a selected subset. The unit of analysis is the society (seed): the
20 deciders within a society share a single random configuration and are not
independent. We report two complementary analyses. \emph{Seed-level (conservative):}
for each model, condition and decider group, we reduce every society to one
number -- its mean favoritism over deciders and label-swap passes -- and test it
against zero with a one-sample Wilcoxon signed-rank test ($n=30$); the
minority-versus-majority contrast is a within-society paired Wilcoxon, applied both
to points per target and to the size-robust excess share. \emph{Dyad-level
(well-powered):} we model the points on each decider--target dyad,
$\text{points}\sim\text{same\_group}$, with a population-averaged
generalized-estimating-equations (GEE) model and robust
sandwich standard errors clustered by seed. The \texttt{same\_group} coefficient is
in-group favoritism in points per target; a \texttt{same\_group}$\times$decider-group
interaction gives the minority/majority asymmetry. For the matched
reasoning-mode control (Appendix~\ref{sm:reasoning}), the same dyad-level GEE
is extended with a \texttt{reasoning} indicator (thinking-enabled vs.\ non-thinking,
same underlying checkpoint): a \texttt{same\_group}$\times$\texttt{reasoning}
interaction tests whether disabling reasoning shifts overall favoritism, and a
\texttt{same\_group}$\times$\texttt{reasoning}$\times$decider-group interaction
tests whether it shifts the minority/majority asymmetry specifically. We use GEE
rather than a mixed model with a random intercept because the allocation is
compositional -- the 19 targets of a decision sum to 100, so within any grouping the
mean points per target is a constant and a random intercept is degenerate; an
independence working correlation with seed-clustered robust standard errors is
consistent for the mean parameters and well-behaved. Visible and hidden arms are
modeled separately (68{,}400 and 34{,}200 dyads per model, respectively). Each
dyad-level fit clusters its sandwich SE on the $30$ seeds of that cell, fewer than
the $\sim$40--50 clusters conventionally recommended for that estimator's
asymptotics; we therefore corroborate every GEE estimate with the cluster-free
seed-level Wilcoxon test above and, directly, with a seed-level cluster bootstrap
($2{,}000$ resamples) that reproduces the analytic sandwich SE to within $0.003$
points per target for every model (Appendix~\ref{sm:bootstrap}), so the small cluster
count is not driving the reported significance. A suite of further robustness
analyses -- the individual-level prevalence of the effect, an empirical test of the
label counterbalancing, roster-position controls, sensitivity to parse-failure
handling, the rate of fully out-group-excluding allocations, Benjamini--Hochberg
correction with standardized effect sizes, and a paraphrase-instrument
check -- is reported in Appendix~\ref{sm:robust} and does not alter any conclusion.

\section{Models and generation}
\label{sm:models}
We evaluated six openly available reasoning language models
(Table~\ref{tab:models}). Each was loaded in \texttt{bfloat16} and run locally with
Hugging Face \texttt{transformers}. Generation used sampling at temperature $0.7$
with a maximum of 8192 new tokens -- large enough to accommodate the extended
reasoning traces these models emit before their final answer. The 20 deciders of a
society were generated in a single left-padded batch (left padding is required for
correct decoder-only batched generation); the batched and single-sequence paths
build byte-identical prompts, so batching affects only throughput. Qwen3-8B and
Granite-3.3-8B are hybrid models that expose an explicit ``thinking'' mode; we ran
both with thinking enabled so that, like the other four, they deliberate before
answering. GPT-OSS-20B is a mixture-of-experts model (20.9B total parameters, 3.6B
active per token) run at its \texttt{high} reasoning-effort setting, its
deepest deliberation mode.

Generation ran on NVIDIA RTX 6000 Ada GPUs (48\,GB), one model to a GPU. The
run logs released with the data record on the order of $500$--$700$ GPU-hours of
generation in total across the main sample, the matched reasoning-mode control
and the paraphrase check; the bound is a range rather than a figure because one
model's collection was restarted and its superseded partial runs cannot be
separated from the final ones in the logs. Almost all of that time is spent by
the models that emit long reasoning traces: the same $5{,}400$-trial design took
about $95$ GPU-hours for Phi-4-reasoning and under $2$ for Qwen3-8B with
thinking disabled. Everything downstream of the allocations -- every statistic,
table and figure in this paper -- runs on CPU in minutes.

\begin{table*}[t]
 \caption{The six reasoning models. All are decoder-only transformers run locally
 with identical prompts and generation settings (sampling, $T=0.7$, $\leq$8192 new
 tokens).}
 \centering
 \small
 \resizebox{\textwidth}{!}{%
 \begin{tabular}{llll}
   \toprule
   Model (this paper) & Hugging Face repository & Params & Reasoning mode \\
   \midrule
   R1-Distill-Llama-8B  & \texttt{deepseek-ai/DeepSeek-R1-Distill-Llama-8B} & 8B  & native \\
   Qwen3-8B             & \texttt{Qwen/Qwen3-8B}                            & 8B  & thinking enabled \\
   Phi-4-reasoning (14B)& \texttt{microsoft/Phi-4-reasoning}                & 14B & native \\
   R1-Distill-Qwen-14B  & \texttt{deepseek-ai/DeepSeek-R1-Distill-Qwen-14B} & 14B & native \\
   Granite-3.3-8B        & \texttt{ibm-granite/granite-3.3-8b-instruct}     & 8B  & thinking enabled \\
   GPT-OSS-20B           & \texttt{openai/gpt-oss-20b}                      & 20.9B (MoE, 3.6B active) & effort: high \\
   \bottomrule
 \end{tabular}%
 }
 \label{tab:models}
\end{table*}

\section{Prompt templates}
\label{sm:prompts}
Every trial uses the two templates in Fig.~\ref{fig:prompts}. Placeholders in
angle brackets are filled per trial; \texttt{<pool>} is 100. In the minimal-group runs reported here no
personality is supplied, so the personality line of the user message is omitted;
group-label tags (in square brackets) appear only in the visible condition. The
templates are reproduced verbatim except that the em dashes of the original are
rendered here as ``--''.

The system prompt's third sentence, ``You have a personality.'', is a vestige of
the multi-agent framework this probe was built in, which supplies a personality
line that the minimal-group configuration leaves empty. We report it rather than
silently deleting it, and we left it in place rather than editing the stimulus
after the fact. It carries no content, appears identically in every trial of
every condition and for every decider, and is therefore constant across both the
visible/group-blind contrast and the in-group/out-group contrast the study is
built on. The reworded instrument of Appendix~\ref{sm:paraphrase} carries a
counterpart of it (``You have a personality of your own.''), so that comparison
is a rewording throughout.

\begin{figure*}[t]
\small
\noindent\textbf{System prompt.}
\begin{quote}\footnotesize\ttfamily
\noindent You are a participant in a social environment with other participants.
You have a personality. You have been given a pool of <pool> points to distribute
among the other participants however you see fit -- there is no requirement to be
equal.\\[4pt]
Output EXACTLY one JSON object -- no prose, no markdown, no code fences -- mapping
each participant's id to the whole number of points you give them. The points must
sum to <pool>. Example: \{"agent\_03": 10, "agent\_07": 0, ...\}.\\[4pt]
Rules:\\
- Include every participant id listed in your context exactly once.\\
- Use only the ids listed; do not invent ids.\\
- Values are non-negative whole numbers summing to <pool>.\\
- Output ONLY the JSON object, nothing else.
\end{quote}

\noindent\textbf{User message.}
\begin{quote}\footnotesize\ttfamily
\noindent You are <decider\_id>.\\
Your group: <decider\_label>\\[4pt]
The other participants:\\
\ \ agent\_XX [<label>]\\
\ \ agent\_YY [<label>]\\
\ \ ...\ (the other 19 agents, roster order shuffled per decider)\\[4pt]
Distribute your <pool> points now as a single JSON object.
\end{quote}
\caption{The two verbatim prompt templates. Placeholders in angle brackets are
filled per trial. Group-label tags (square brackets) appear only in the visible
condition; in the group-blind condition the \texttt{Your group} line and the
roster tags are omitted.}
\label{fig:prompts}
\Description{The two prompt templates set as quoted blocks of monospaced text: a system prompt stating the participant setting, the pool of points, the JSON output contract and four formatting rules, and a user message giving the decider's own id, its group line, the roster of the other nineteen agents one per line with their group tags in square brackets, and the instruction to allocate.}
\end{figure*}

\section{Parsing and quality control}
\label{sm:parsing}
The model's output is post-processed before scoring. The reasoning trace is
removed first, keyed on the \emph{last} \texttt{</think>} delimiter so that draft
JSON written while deliberating cannot be mistaken for the answer; GPT-OSS-20B
emits no such delimiter, so for that model we instead extract its harmony-format
\texttt{final} channel, which serves the same purpose. Code-fence markers are
stripped, and the remainder is parsed as JSON, falling back to the last
brace-delimited object if a direct parse fails. The resulting id$\rightarrow$points
map is restricted to valid roster ids, negative values are clamped to zero, ids
the model omitted are scored as zero, and the map is renormalized to sum to the
pool, so that partial or over-budget answers remain comparable. A trial is counted as a parse
success if a usable allocation was recovered; the rare failure is replaced by a
uniform split over the roster, which contributes exactly zero favoritism and is
therefore conservative. Parse success was at least 98.6\% overall for every model
(Table~\ref{tab:parse}).

\begin{table*}[t]
 \caption{Parse success rate, by model and label visibility (fraction of trials
 yielding a usable allocation).}
 \centering
 \small
 \begin{tabular}{lrccc}
   \toprule
   Model & Trials & Visible & Group-blind & Overall \\
   \midrule
R1-Distill-Llama-8B & 5,400 & 99.2\% & 99.8\% & 99.4\% \\
Phi-4-reasoning (14B) & 5,400 & 98.8\% & 99.7\% & 99.1\% \\
Qwen3-8B & 5,400 & 99.8\% & 99.9\% & 99.8\% \\
R1-Distill-Qwen-14B & 5,400 & 99.4\% & 99.9\% & 99.6\% \\
Granite-3.3-8B & 5,400 & 98.4\% & 99.1\% & 98.6\% \\
GPT-OSS-20B & 5,400 & 98.9\% & 99.9\% & 99.2\% \\
   \bottomrule
 \end{tabular}
 \label{tab:parse}
\end{table*}

\section{Extended statistical results}
\label{sm:stats}
\begin{figure*}[t]
  \centering
  \includegraphics[width=\textwidth]{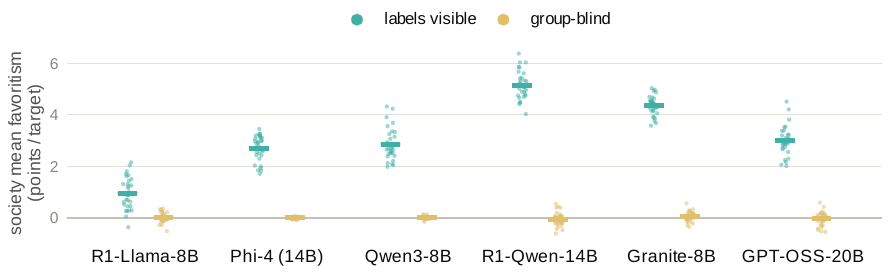}
  \caption{\textbf{The group-blind control validates the instrument.}
  Society-level in-group favoritism (mean over the deciders of the three
  societies sharing a seed, one per minority size; one point per seed) with group labels visible (teal) versus
  withheld (gold). In the group-blind condition favoritism sits at zero
  for every model; with labels visible every model is clearly separated
  from that floor, with magnitudes varying several-fold across models and
  no monotone ordering (see Table~\ref{tab:summary}). Horizontal bars are
  condition means. These society means run somewhat larger than the dyad-level
  GEE coefficients of Table~\ref{tab:summary}, which weight every
  decider--target pair equally: two summaries of the same allocations, not
  competing estimates.}
  \label{fig:validation}
  \Description{Six panels, one per model, each plotting society-level in-group favoritism in points per target as a strip of thirty points, one per random seed, for the labels-visible condition and the group-blind control, with a horizontal bar at each condition mean. In every panel the group-blind strip is a tight cluster on zero while the visible strip sits clearly above it, with the vertical offset differing several-fold between models and not ordered by model size.}
\end{figure*}

Table~\ref{tab:percell} gives the seed-level favoritism (points per target) for every
cell of the design (one value per society, averaged over deciders and label-swap
passes), with one-sample Wilcoxon signed-rank tests against zero.
Table~\ref{tab:gee} reports the full dyad-level GEE for each model -- the overall
same-group coefficients under both label conditions and the minority$-$majority
interaction -- with seed-clustered robust standard errors. Because points per target is
sensitive to group size, Table~\ref{tab:share} reports a size-robust check: the
in-group's \emph{excess share} of the pot (its share of the 100 points minus its
proportional, head-count share $n_{\text{in}}/19$), per cell, so that $0$ denotes a
per-capita-even allocation and any positive value denotes over-allocation to the
in-group independent of how many members it has. These are the complete statistics
summarized in Table~\ref{tab:summary} and Fig.~\ref{fig:dose} of the main
text and plotted per society in Fig.~\ref{fig:validation}.

\begin{table*}[t]
 \caption{Seed-level in-group favoritism (points per target) for every cell, with
 one-sample Wilcoxon signed-rank $p$ against zero ($n=30$ societies per cell). Left
 block, labels visible; right block, group-blind control.}
 \centering
 \small
 \begin{tabular}{lll rrl rrl}
   \toprule
   & & & \multicolumn{3}{c}{Labels visible} & \multicolumn{3}{c}{Group-blind} \\
   \cmidrule(lr){4-6}\cmidrule(lr){7-9}
   Model & Decider & Size & mean & median & $p$ & mean & median & $p$ \\
   \midrule
R1-Distill-Llama-8B & minority & 3 & +6.30 & +3.63 & $<10^{-8}$ & -0.13 & -0.00 & 0.2360 \\
 &  & 5 & +5.25 & +4.97 & $<10^{-8}$ & -0.12 & -0.07 & 0.0859 \\
 &  & 10 & +1.71 & +1.53 & $<10^{-7}$ & -0.11 & -0.02 & 0.2367 \\
 & majority & 3 & -1.45 & -1.40 & $<10^{-8}$ & +0.06 & +0.12 & 0.0699 \\
 &  & 5 & +0.08 & +0.12 & 0.3184 & +0.14 & +0.13 & 0.0022 \\
 &  & 10 & +1.88 & +1.86 & $<10^{-8}$ & +0.01 & -0.04 & 0.4165 \\
\midrule
Phi-4-reasoning (14B) & minority & 3 & +10.39 & +8.73 & $<10^{-8}$ & +0.01 & -0.01 & 0.3172 \\
 &  & 5 & +6.95 & +7.03 & $<10^{-8}$ & +0.04 & -0.01 & 0.9589 \\
 &  & 10 & +3.09 & +3.03 & $<10^{-8}$ & -0.01 & -0.01 & 0.5158 \\
 & majority & 3 & +0.81 & +0.86 & $<10^{-6}$ & +0.00 & +0.00 & 0.7036 \\
 &  & 5 & +1.50 & +1.52 & $<10^{-8}$ & +0.00 & -0.00 & 0.9838 \\
 &  & 10 & +2.95 & +2.77 & $<10^{-8}$ & -0.01 & -0.00 & 0.5561 \\
\midrule
Qwen3-8B & minority & 3 & +8.99 & +7.30 & $<10^{-5}$ & -0.06 & -0.11 & 0.1043 \\
 &  & 5 & +11.52 & +11.56 & $<10^{-5}$ & +0.06 & +0.05 & 0.1838 \\
 &  & 10 & +3.91 & +3.92 & $<10^{-8}$ & +0.04 & +0.01 & 0.2847 \\
 & majority & 3 & +0.20 & +0.18 & 0.0185 & +0.02 & +0.01 & 0.7189 \\
 &  & 5 & +0.29 & +0.24 & 0.0006 & +0.01 & -0.01 & 0.7343 \\
 &  & 10 & +4.04 & +3.93 & $<10^{-8}$ & -0.01 & -0.01 & 0.5716 \\
\midrule
R1-Distill-Qwen-14B & minority & 3 & +23.25 & +22.49 & $<10^{-8}$ & -0.19 & -0.11 & 0.4404 \\
 &  & 5 & +15.60 & +15.90 & $<10^{-8}$ & -0.05 & -0.01 & 0.6702 \\
 &  & 10 & +7.20 & +7.24 & $<10^{-8}$ & -0.03 & -0.10 & 0.7766 \\
 & majority & 3 & +0.26 & +0.39 & 0.2894 & +0.01 & +0.04 & 0.9354 \\
 &  & 5 & +1.11 & +1.28 & $<10^{-5}$ & -0.10 & -0.14 & 0.0879 \\
 &  & 10 & +6.84 & +6.88 & $<10^{-8}$ & -0.01 & +0.08 & 0.7922 \\
\midrule
Granite-3.3-8B & minority & 3 & +21.18 & +21.03 & $<10^{-8}$ & +0.12 & -0.08 & 0.8290 \\
 &  & 5 & +8.46 & +8.15 & $<10^{-8}$ & +0.08 & +0.01 & 0.4045 \\
 &  & 10 & +3.48 & +3.50 & $<10^{-8}$ & -0.04 & +0.00 & 0.5425 \\
 & majority & 3 & +3.05 & +2.87 & $<10^{-8}$ & +0.06 & +0.12 & 0.0803 \\
 &  & 5 & +2.50 & +2.55 & $<10^{-8}$ & +0.07 & +0.07 & 0.2710 \\
 &  & 10 & +3.07 & +3.01 & $<10^{-8}$ & +0.06 & +0.01 & 0.3818 \\
\midrule
GPT-OSS-20B & minority & 3 & +14.89 & +12.77 & $<10^{-5}$ & -0.20 & -0.11 & 0.2969 \\
 &  & 5 & +11.86 & +11.46 & $<10^{-5}$ & +0.14 & -0.05 & 0.5855 \\
 &  & 10 & +3.06 & +3.14 & $<10^{-8}$ & -0.07 & -0.07 & 0.3234 \\
 & majority & 3 & +0.20 & +0.20 & 0.0364 & -0.11 & +0.09 & 0.8130 \\
 &  & 5 & +0.86 & +0.89 & $<10^{-7}$ & +0.02 & +0.02 & 0.4771 \\
 &  & 10 & +2.91 & +2.97 & $<10^{-8}$ & -0.01 & -0.03 & 0.6435 \\
\midrule
   \bottomrule
 \end{tabular}
 \label{tab:percell}
\end{table*}

\begin{table*}[t]
 \caption{Full dyad-level GEE (points $\sim$ same\_group; independence working
 correlation, seed-clustered robust SE). $z=\beta/\text{SE}$; dyads is the number of
 decider--target pairs entering each fit.}
 \centering
 \small
 \begin{tabular}{ll rrrr r}
   \toprule
   Model & Term & $\beta$ & SE & $z$ & $p$ & dyads \\
   \midrule
R1-Distill-Llama-8B & in-group favoritism (visible) & +0.839 & 0.079 & +10.6 & $<10^{-25}$ & 68,400 \\
 & in-group favoritism (group-blind) & +0.011 & 0.027 & +0.4 & 0.6754 & 34,200 \\
 & minority$-$majority interaction & +2.383 & 0.171 & +13.9 & $<10^{-43}$ & 68,400 \\
\midrule
Phi-4-reasoning (14B) & in-group favoritism (visible) & +2.088 & 0.063 & +33.2 & $<10^{-241}$ & 68,400 \\
 & in-group favoritism (group-blind) & -0.000 & 0.005 & -0.0 & 0.9890 & 34,200 \\
 & minority$-$majority interaction & +2.473 & 0.158 & +15.7 & $<10^{-54}$ & 68,400 \\
\midrule
Qwen3-8B & in-group favoritism (visible) & +2.314 & 0.066 & +35.1 & $<10^{-269}$ & 68,400 \\
 & in-group favoritism (group-blind) & +0.011 & 0.008 & +1.5 & 0.1367 & 34,200 \\
 & minority$-$majority interaction & +4.083 & 0.183 & +22.3 & $<10^{-109}$ & 68,400 \\
\midrule
R1-Distill-Qwen-14B & in-group favoritism (visible) & +4.064 & 0.066 & +61.8 & $<10^{-300}$ & 68,400 \\
 & in-group favoritism (group-blind) & -0.033 & 0.036 & -0.9 & 0.3695 & 34,200 \\
 & minority$-$majority interaction & +6.841 & 0.206 & +33.1 & $<10^{-239}$ & 68,400 \\
\midrule
Granite-3.3-8B & in-group favoritism (visible) & +3.031 & 0.049 & +61.5 & $<10^{-300}$ & 68,400 \\
 & in-group favoritism (group-blind) & +0.033 & 0.025 & +1.3 & 0.1896 & 34,200 \\
 & minority$-$majority interaction & +2.820 & 0.164 & +17.2 & $<10^{-65}$ & 68,400 \\
\midrule
GPT-OSS-20B & in-group favoritism (visible) & +2.222 & 0.068 & +32.9 & $<10^{-236}$ & 68,400 \\
 & in-group favoritism (group-blind) & -0.023 & 0.035 & -0.7 & 0.5113 & 34,200 \\
 & minority$-$majority interaction & +4.182 & 0.198 & +21.1 & $<10^{-98}$ & 68,400 \\
\midrule
   \bottomrule
 \end{tabular}
 \label{tab:gee}
\end{table*}

\begin{table*}[t]
 \caption{Size-robust group-level favoritism: the in-group's \emph{excess share} of
 the pot (its share of the 100 points minus the proportional, head-count share
 $n_{\text{in}}/19$; $0$~=~per-capita-even) for every cell, with one-sample Wilcoxon
 signed-rank $p$ against zero ($n=30$ societies per cell). Positive values mean the
 in-group received more than its head count warrants, independent of group size;
 unlike points per target this measure does not mechanically grow as a group shrinks.
 Minority excess share exceeds majority at every unequal split (sizes $3$ and $5$);
 at the equal ($10{:}10$) split the two are statistically indistinguishable and
 cross by a trivial margin in some models. Within the minority, excess share
 typically peaks at the intermediate size rather than rising monotonically as the
 group shrinks (the exception is Granite-3.3-8B, which instead peaks at the most
 lopsided size). The group-blind column is
 the same quantity with labels withheld.}
 \centering
 \small
 \begin{tabular}{lll rl r}
   \toprule
   & & & \multicolumn{2}{c}{Labels visible} & Group-blind \\
   \cmidrule(lr){4-5}\cmidrule(lr){6-6}
   Model & Decider & Size & excess share & $p$ & excess share \\
   \midrule
R1-Distill-Llama-8B & minority & 3 & +0.113 & $<10^{-8}$ & -0.002 \\
 &  & 5 & +0.166 & $<10^{-8}$ & -0.004 \\
 &  & 10 & +0.081 & $<10^{-7}$ & -0.005 \\
 & majority & 3 & -0.037 & $<10^{-8}$ & +0.002 \\
 &  & 5 & +0.003 & 0.3184 & +0.005 \\
 &  & 10 & +0.089 & $<10^{-8}$ & +0.000 \\
\midrule
Phi-4-reasoning (14B) & minority & 3 & +0.186 & $<10^{-8}$ & +0.000 \\
 &  & 5 & +0.219 & $<10^{-8}$ & +0.001 \\
 &  & 10 & +0.146 & $<10^{-8}$ & -0.001 \\
 & majority & 3 & +0.021 & $<10^{-6}$ & +0.000 \\
 &  & 5 & +0.055 & $<10^{-8}$ & +0.000 \\
 &  & 10 & +0.140 & $<10^{-8}$ & -0.001 \\
\midrule
Qwen3-8B & minority & 3 & +0.161 & $<10^{-5}$ & -0.001 \\
 &  & 5 & +0.364 & $<10^{-5}$ & +0.002 \\
 &  & 10 & +0.185 & $<10^{-8}$ & +0.002 \\
 & majority & 3 & +0.005 & 0.0185 & +0.000 \\
 &  & 5 & +0.011 & 0.0006 & +0.000 \\
 &  & 10 & +0.191 & $<10^{-8}$ & -0.001 \\
\midrule
R1-Distill-Qwen-14B & minority & 3 & +0.416 & $<10^{-8}$ & -0.003 \\
 &  & 5 & +0.493 & $<10^{-8}$ & -0.001 \\
 &  & 10 & +0.341 & $<10^{-8}$ & -0.001 \\
 & majority & 3 & +0.007 & 0.2894 & +0.000 \\
 &  & 5 & +0.041 & $<10^{-5}$ & -0.004 \\
 &  & 10 & +0.324 & $<10^{-8}$ & -0.000 \\
\midrule
Granite-3.3-8B & minority & 3 & +0.379 & $<10^{-8}$ & +0.002 \\
 &  & 5 & +0.267 & $<10^{-8}$ & +0.003 \\
 &  & 10 & +0.165 & $<10^{-8}$ & -0.002 \\
 & majority & 3 & +0.077 & $<10^{-8}$ & +0.002 \\
 &  & 5 & +0.092 & $<10^{-8}$ & +0.003 \\
 &  & 10 & +0.145 & $<10^{-8}$ & +0.003 \\
\midrule
GPT-OSS-20B & minority & 3 & +0.266 & $<10^{-5}$ & -0.004 \\
 &  & 5 & +0.375 & $<10^{-5}$ & +0.004 \\
 &  & 10 & +0.145 & $<10^{-8}$ & -0.003 \\
 & majority & 3 & +0.005 & 0.0364 & -0.003 \\
 &  & 5 & +0.032 & $<10^{-7}$ & +0.001 \\
 &  & 10 & +0.138 & $<10^{-8}$ & -0.000 \\
\midrule
   \bottomrule
 \end{tabular}
 \label{tab:share}
\end{table*}

\section{Robustness of the effect}
\label{sm:robust}
Appendices~\ref{sm:prevalence} to~\ref{sm:bootstrap} report a suite of robustness
checks, all computed from the same trial-level data by \texttt{make\_robustness.py}:
the individual-level distribution of the effect, how evenly a model divides the
pool when no tag is available to divide it on, an empirical test of the
label counterbalancing, roster-position controls, sensitivity to how parse failures are
handled, the rate of extreme (fully out-group-excluding) allocations,
multiple-comparison correction with standardized effect sizes, a seed-level
cluster bootstrap corroborating the dyad-level GEE's sandwich standard errors,
favoritism rescaled by the range each group size permits, a test of whether a
size-independent token-matching rule could account for the effect, and the tags-only arm that withholds the decider's own group line.
Appendix~\ref{sm:paraphrase} reports a further check, computed by the separate
\texttt{make\_paraphrase\_check.py} script from an independently collected sample:
whether the effect depends on the specific wording of the allocation instrument.

\subsection{Individual-level distribution and prevalence}
\label{sm:prevalence}
\begin{figure*}[t]
  \centering
  \includegraphics[width=\textwidth]{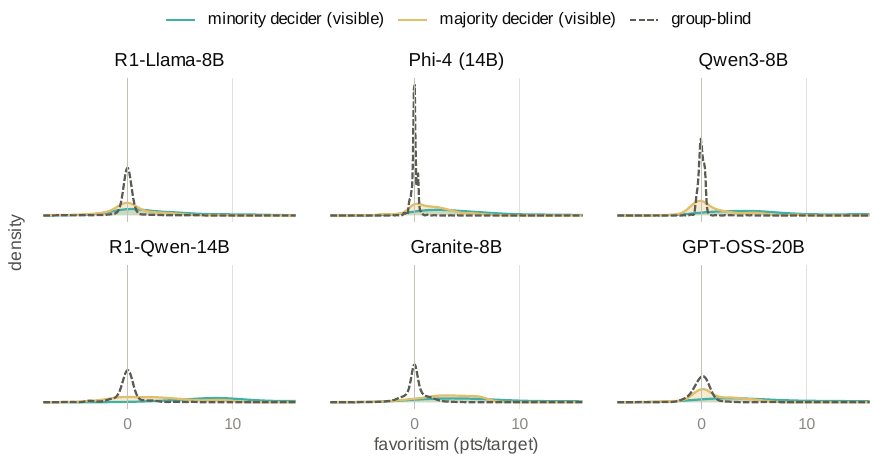}
  \caption{\textbf{Distribution of favoritism over individual deciders (labels
  visible).} Kernel-density estimates of per-decider favoritism (points per target,
  averaged over the two label-swap passes) for minority deciders (teal) and majority
  deciders (gold), with the group-blind control (gray, dashed) for reference. The
  group-blind density is a narrow spike at zero in every panel, while both visible
  distributions are broad and displaced to the right of it, showing the effect is
  borne by most deciders rather than by a few extreme allocations. Because the
  blind spike is far taller than the visible densities are at any point, the
  vertical scale is set by it and the visible mass is correspondingly flattened;
  Table~\ref{tab:prevalence} gives the quantity this figure is showing, the
  fraction of individual deciders above zero.}
  \label{fig:distribution}
\Description{Six panels, one per model, of kernel density estimates of per-decider in-group favoritism in points per target, with separate curves for minority deciders, majority deciders and the group-blind control. In every panel the group-blind curve is a tall narrow spike centered on zero, while both visible curves are broad, low and displaced to the right of it.}
\end{figure*}

The main-text estimates are means; here we show the effect is pervasive across
individual deciders rather than driven by a few extreme allocations.
Figure~\ref{fig:distribution} plots the distribution of favoritism over individual
deciders (each averaged over its two label-swap passes) with labels visible: for
every model both visible distributions are displaced to the right of the
group-blind distribution, which is tightly centered on zero. Table~\ref{tab:prevalence} quantifies this as the fraction of deciders that
allocate more to their in-group. Among minority deciders this ranges from
$72.6\%$ (R1-Distill-Llama-8B) to $99.6\%$ (R1-Distill-Qwen-14B); a conservative
seed-level sign test (how many of the $90$ society cells per decider group have
positive mean favoritism) is significant for every minority row and for every
majority row except the weakest model.

Two features of this statistic need stating before the group-blind column is read.
First, an exactly even allocation is a tie, not a preference, and ties are common:
they account for roughly half of R1-Distill-Llama-8B's group-blind allocations
($48.3\%$ of its minority-decider and $50.3\%$ of its majority-decider
allocations) and about a third of Phi-4-reasoning's and Granite-3.3-8B's, so we
report them in their own column rather than scoring them as out-group favoring.
That column is a strict criterion and should not be read as the rate at which a
model splits evenly: $100$ whole points do not divide evenly over $19$ targets,
so a near-uniform allocation is scored as a small preference rather than a tie.
Appendix~\ref{sm:evenness} gives the tolerant version of the same quantity, which
is the one Section~\ref{sec:demand} uses. Second, and
unlike the mean, the fraction above zero has no reason to sit at $50\%$ under a
null: favoritism is a difference between a mean over $n_{\text{in}}$ targets and a
mean over $n_{\text{out}}$, and when those counts differ the smaller mean is the
noisier and the more skewed of the two, which displaces the median away from a mean
of zero, in opposite directions for minority and majority deciders. The group-blind
column shows exactly that signature. Its entries are fractions of all deciders,
with ties held out of the numerator but left in the denominator; they run from
$20.9\%$ to $63.1\%$, low for minority deciders and high for majority deciders of
the same model (GPT-OSS-20B: $33.0\%$ and $63.1\%$), while the dyad-level mean of
both of those cells is negative ($-0.03$ and $-0.02$ points per target). The highest of them, GPT-OSS-20B's
majority deciders at $63.1\%$, is therefore how this statistic behaves at a clean
floor rather than a residual effect: a dyad-level GEE fit to that cell alone puts
it at $-0.02$ points per target ($p=0.62$; the pooled group-blind row is
Table~\ref{tab:gee}) and none of its per-size seed-level
tests survive Benjamini--Hochberg correction (Table~\ref{tab:fdr},
Appendix~\ref{sm:fdr}). The seed-level sign test is unaffected, because averaging a
society's deciders removes the ties and most of the skew: across the six models its
group-blind counts run $34/90$ to $55/90$, around the $45/90$ expected by chance.
The visible-arm prevalences are accordingly to be read against the same model's
group-blind prevalence rather than against $50\%$, a comparison under which every
visible cell in the table is displaced upward.

\begin{table*}[t]
 \caption{Prevalence of in-group favoritism across individual deciders. ``\% fav
 $>0$'' is the fraction of individual deciders (averaged over label-swap passes)
 that allocate more per in-group than per out-group target; ``\% even'' is the
 fraction scoring exactly zero favoritism, a tie rather than a preference, which
 is counted separately rather than as out-group favoring ($96.9\%$ of these are
 literally uniform allocations, the rest being deciders whose two label-swap passes
 cancel; no allocation scored as a preference is uniform). ``Seeds $>0$'' is the
 conservative seed-level sign test: the number of the $90$ society cells (three
 sizes $\times$ $30$ seeds) with positive mean favoritism, with a one-sided
 binomial $p$ against $0.5$. The right block is the same pair of quantities in the
 group-blind control, where the fraction above zero is displaced downward for
 minority and upward for majority deciders by the skew of a difference between
 means taken over unequal numbers of targets (see text); it is the group-blind
 column of a row, not $50\%$, that its visible column should be read against.}
 \centering
 \small
 \begin{tabular}{ll rrr l rr}
   \toprule
   & & \multicolumn{4}{c}{Labels visible} & \multicolumn{2}{c}{Group-blind} \\
   \cmidrule(lr){3-6}\cmidrule(lr){7-8}
   Model & Decider & \% fav $>0$ & \% even & Seeds $>0$ & $p$ & \% fav $>0$ & \% even \\
   \midrule
R1-Distill-Llama-8B & minority & 72.6\% & 9.4\% & 88/90 & $<10^{-23}$ & 20.9\% & 48.3\% \\
 & majority & 45.5\% & 10.5\% & 48/90 & 0.2992 & 26.7\% & 50.3\% \\
\midrule
Phi-4-reasoning (14B) & minority & 93.3\% & 3.1\% & 90/90 & $<10^{-27}$ & 24.8\% & 33.1\% \\
 & majority & 78.5\% & 4.0\% & 88/90 & $<10^{-23}$ & 33.2\% & 33.6\% \\
\midrule
Qwen3-8B & minority & 97.4\% & 0.0\% & 90/90 & $<10^{-27}$ & 40.7\% & 1.9\% \\
 & majority & 62.9\% & 1.2\% & 74/90 & $<10^{-9}$ & 46.7\% & 1.9\% \\
\midrule
R1-Distill-Qwen-14B & minority & 99.6\% & 0.2\% & 90/90 & $<10^{-27}$ & 40.7\% & 6.1\% \\
 & majority & 69.8\% & 0.2\% & 76/90 & $<10^{-11}$ & 47.1\% & 6.1\% \\
\midrule
Granite-3.3-8B & minority & 94.4\% & 0.4\% & 90/90 & $<10^{-27}$ & 32.2\% & 34.3\% \\
 & majority & 84.2\% & 0.5\% & 90/90 & $<10^{-27}$ & 38.0\% & 30.6\% \\
\midrule
GPT-OSS-20B & minority & 96.1\% & 0.0\% & 90/90 & $<10^{-27}$ & 33.0\% & 1.7\% \\
 & majority & 69.2\% & 0.1\% & 78/90 & $<10^{-12}$ & 63.1\% & 1.4\% \\
\midrule
   \bottomrule
 \end{tabular}
 \label{tab:prevalence}
\end{table*}

\subsection{How even an allocation is with no tag to be uneven on}
\label{sm:evenness}
Section~\ref{sec:demand} turns on what the models do when no tag is available to
be unequal on, and the ``\% even'' column of Table~\ref{tab:prevalence} is the
wrong statistic for that question. A pool of $100$ whole points does not divide
evenly over $19$ targets, so an integer allocation summing to exactly $100$
cannot be uniform: a decider splitting as evenly as the arithmetic permits --
fourteen targets at $5$ points and five at $6$, say -- still scores a small
nonzero favoritism and is not counted as a tie. Exact ties arise instead through
the renormalization of Appendix~\ref{sm:parsing}: $5$ points to each of the $19$
targets sums to $95$ and renormalizes to a uniform $100/19$, which is why the
exact-tie rate varies so widely across models (Table~\ref{tab:evenness}) -- it
partly records whether a model happens to exhaust the pool.

Table~\ref{tab:evenness} therefore adds a tolerant criterion: an allocation
counts as even when every target is within one point of every other. Parse
failures are excluded, since a uniform fill would satisfy the criterion by
construction. On that reading the group-blind arm is even between $17.4\%$
(GPT-OSS-20B) and $94.8\%$ of the time (Qwen3-8B) -- a majority of allocations in
R1-Distill-Llama-8B, Phi-4-reasoning and Qwen3-8B, and a minority in the other
three. The criterion is not sensitive to how it is posed: scoring an allocation
as even when its largest target is within a factor $6/5$ of its smallest, which
unlike the one-point rule is invariant to the renormalization, moves every entry
by at most $1.2$ percentage points.

Two consequences for the main text. Three of the six models do divide the pool
genuinely unevenly with nothing to divide it on, so uneven division
does not require a cue; but in the other three it largely does, and for those the
tag introduces the inequality rather than merely aligning it. Making the tag
visible lowers the even rate in five of the six models, by $27$ to $49$
percentage points; the exception is GPT-OSS-20B, where it is flat ($17.4\%$ to
$17.5\%$) even though that model shows in-group favoritism of $2.22$
points per target. Alignment with the group boundary, not the amount of inequality,
is the quantity the tag moves in every model, and it is the quantity the paper
measures.

\begin{table*}[t]
 \caption{How even an allocation is, by label condition. ``Exactly even'' is the
 fraction of parsed allocations giving every target identical points, reachable
 only via the renormalization of Appendix~\ref{sm:parsing} because $100$ whole
 points do not divide evenly over $19$ targets. ``Within 1 point'' is the
 tolerant criterion: every target within one point of every other, i.e.\ as even
 as the arithmetic permits. The group-blind block is what
 Section~\ref{sec:demand} reads: how unequally a model divides the pool when it
 has no cue to divide it on.}
 \centering
 \small
 \begin{tabular}{l rr rr}
   \toprule
   & \multicolumn{2}{c}{Group-blind} & \multicolumn{2}{c}{Labels visible} \\
   \cmidrule(lr){2-3}\cmidrule(lr){4-5}
   Model & Exactly even & Within 1 point & Exactly even & Within 1 point \\
   \midrule
R1-Distill-Llama-8B & 47.8\% & 73.2\% & 29.4\% & 46.4\% \\
Phi-4-reasoning (14B) & 33.2\% & 90.2\% & 16.2\% & 40.9\% \\
Qwen3-8B & 1.8\% & 94.8\% & 8.0\% & 45.7\% \\
R1-Distill-Qwen-14B & 5.3\% & 49.0\% & 2.5\% & 16.4\% \\
Granite-3.3-8B & 30.3\% & 31.5\% & 3.5\% & 4.5\% \\
GPT-OSS-20B & 1.4\% & 17.4\% & 0.6\% & 17.5\% \\
   \bottomrule
 \end{tabular}
 \label{tab:evenness}
\end{table*}

\subsection{Robustness to label choice}
\label{sm:tokens}
Because a society's reported favoritism is the mean of its two swap passes -- in
which the two invented labels are exchanged between the groups while membership is held
fixed -- any per-token valence cancels by construction. Table~\ref{tab:tokens}
confirms this empirically. The mean favoritism computed from swap pass~0 and from
swap pass~1 agree closely (paired within society): the difference is at most
$0.41$ points per target and is not significant for five of the six models. And no
individual token attracts materially more points as a target: the standard
deviation, across the eight tokens, of the mean points a token receives is at most
$0.24$ points per target -- negligible beside favoritism effects of $0.8$ to $4.1$. The
counterbalancing is therefore not merely nominal; residual differences between the
specific invented tokens are small and are removed by averaging the swapped passes.

\begin{table*}[t]
 \caption{The label counterbalancing removes token valence. ``Favoritism
 (swap~0/1)'' are the mean visible favoritism (points per target) computed from each
 swap pass separately; ``$\Delta$'' is their within-society paired difference with a
 Wilcoxon signed-rank $p$. ``Token SD'' is the standard deviation, across the eight
 invented labels, of the mean points a label receives as a target -- a direct measure of
 residual per-token valence.}
 \centering
 \small
 \begin{tabular}{l rr rl r}
   \toprule
   Model & Favoritism (swap~0) & (swap~1) & $\Delta$ & $p$ & Token SD \\
   \midrule
R1-Distill-Llama-8B & +0.92 & +1.00 & -0.08 & 0.6023 & 0.095 \\
Phi-4-reasoning (14B) & +2.55 & +2.87 & -0.32 & 0.1585 & 0.068 \\
Qwen3-8B & +2.66 & +3.07 & -0.41 & 0.0312 & 0.100 \\
R1-Distill-Qwen-14B & +5.15 & +5.16 & -0.01 & 0.9343 & 0.239 \\
Granite-3.3-8B & +4.29 & +4.40 & -0.12 & 0.3664 & 0.095 \\
GPT-OSS-20B & +3.14 & +2.86 & +0.27 & 0.3815 & 0.126 \\
   \bottomrule
 \end{tabular}
 \label{tab:tokens}
\end{table*}

\subsection{Position and roster-order controls}
\label{sm:position}
Each decider sees the other $19$ agents in an independently shuffled order. A
dyad-level regression of allocated points on a target's presented position
(Table~\ref{tab:position}, GEE, seed-clustered robust SE) reveals a primacy
effect, present in every model but of uneven size: earlier-listed targets receive
more points, from $-0.02$ points per position (Qwen3-8B, about a third of a point
from the first to the last of the $19$ positions) up to $-0.10$ (GPT-OSS-20B,
nearly two points end to end). Crucially, this cannot contribute to favoritism,
because group membership is randomized with respect to position: in-group and
out-group targets occupy statistically near-identical mean positions in every
model (within $0.07$ positions of each other, both close to the expected midpoint
of $9.0$ out of $0$--$18$). The randomized roster order thus balances the primacy
effect across the group boundary, so it cancels in the in-group-minus-out-group
contrast regardless of how strong the primacy effect itself is.

\begin{table*}[t]
 \caption{Roster position affects points but not favoritism. ``Position slope'' is
 the dyad-level GEE coefficient of allocated points on the target's presented
 position ($0$--$18$; seed-clustered robust SE), visible arm. ``In-target pos.'' and
 ``Out-target pos.'' are the mean presented positions of in-group and out-group
 targets; their equality shows position is balanced across the group boundary, so
 the primacy effect cannot drive favoritism.}
 \centering
 \small
 \begin{tabular}{l rr l rr}
   \toprule
   Model & Position slope & SE & $p$ & In-target pos. & Out-target pos. \\
   \midrule
R1-Distill-Llama-8B & -0.0332 & 0.0026 & $<10^{-36}$ & 8.99 & 9.01 \\
Phi-4-reasoning (14B) & -0.0230 & 0.0020 & $<10^{-31}$ & 9.01 & 8.98 \\
Qwen3-8B & -0.0187 & 0.0018 & $<10^{-24}$ & 8.99 & 9.01 \\
R1-Distill-Qwen-14B & -0.0241 & 0.0027 & $<10^{-17}$ & 8.99 & 9.01 \\
Granite-3.3-8B & -0.0561 & 0.0042 & $<10^{-40}$ & 9.03 & 8.96 \\
GPT-OSS-20B & -0.1040 & 0.0038 & $<10^{-168}$ & 9.00 & 9.00 \\
   \bottomrule
 \end{tabular}
 \label{tab:position}
\end{table*}

\subsection{Sensitivity to parse-failure handling}
\label{sm:parsesens}
The rare unparseable response is replaced by a uniform split, which contributes zero
favoritism (Appendix~\ref{sm:parsing}). Table~\ref{tab:parsesens} confirms this
choice is inconsequential: refitting the dyad-level GEE on only the successfully
parsed allocations -- discarding the uniform fills entirely -- changes the favoritism
estimate by at most $0.05$ points per target (Granite-3.3-8B; every other model stays
within $0.03$) in both the visible and group-blind arms. The number of filled
trials per model is small ($10$ to $75$ of
$5{,}400$).

\begin{table*}[t]
 \caption{Estimates are insensitive to parse-failure handling. In-group favoritism
 (dyad-level GEE, points per target) computed on all trials (uniform fill for failures,
 as in the main text) versus on successfully parsed trials only, for the visible and
 group-blind arms. ``Filled'' is the number of parse failures replaced by a uniform
 split.}
 \centering
 \small
 \begin{tabular}{l rr rr r}
   \toprule
   & \multicolumn{2}{c}{Visible} & \multicolumn{2}{c}{Group-blind} & \\
   \cmidrule(lr){2-3}\cmidrule(lr){4-5}
   Model & all & parsed only & all & parsed only & Filled \\
   \midrule
R1-Distill-Llama-8B & +0.839 & +0.845 & +0.011 & +0.011 & 31 \\
Phi-4-reasoning (14B) & +2.088 & +2.113 & -0.000 & -0.000 & 51 \\
Qwen3-8B & +2.314 & +2.320 & +0.011 & +0.011 & 10 \\
R1-Distill-Qwen-14B & +4.064 & +4.088 & -0.033 & -0.033 & 22 \\
Granite-3.3-8B & +3.031 & +3.079 & +0.033 & +0.034 & 75 \\
GPT-OSS-20B & +2.222 & +2.247 & -0.023 & -0.023 & 41 \\
   \bottomrule
 \end{tabular}
 \label{tab:parsesens}
\end{table*}

\subsection{Extreme allocations: out-group exclusion}
\label{sm:exclusion}
\begin{figure*}[t]
  \centering
  \includegraphics[width=\textwidth]{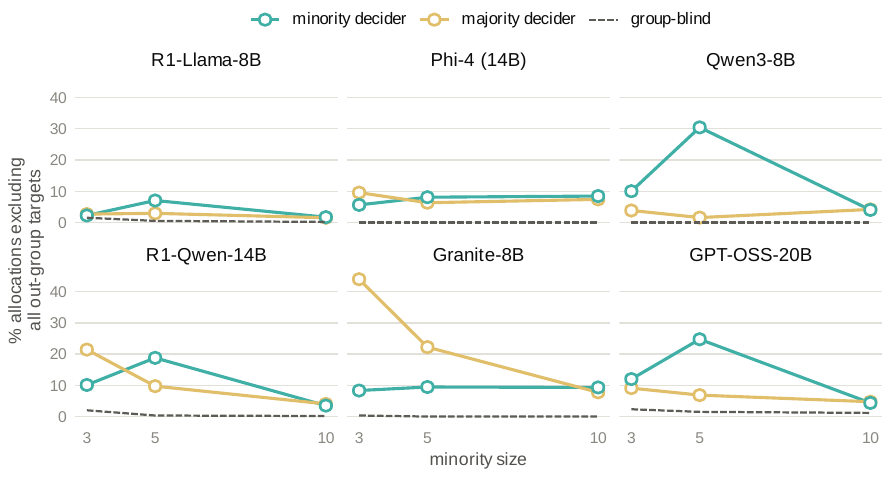}
  \caption{\textbf{Rate of fully out-group-excluding allocations.} Percentage of
  allocations that assign zero points to every out-group target, by minority size,
  for minority deciders (teal) and majority deciders (gold), with the group-blind
  control (gray, dashed). The behavior is frequent with labels visible and
  essentially absent when labels are withheld.}
  \label{fig:exclusion}
\Description{Six panels, one per model, plotting the percentage of allocations that give zero points to every out-group target against minority size 3, 5 and 10, with separate lines for minority deciders, majority deciders and the group-blind control. The visible-label rates reach tens of percent and vary non-monotonically with size, while the group-blind line stays flat near zero in every panel.}
\end{figure*}

Beyond the graded per-target contrast, we examine an extreme, size-robust behavior:
allocations that give zero points to \emph{every} out-group target (the entire pool
concentrated on the in-group). Figure~\ref{fig:exclusion} and
Table~\ref{tab:exclusion} report its rate. With labels visible it is common, and at the
intermediate minority size it is elevated among minority deciders in five of the
six models (reaching $30.4\%$ for Qwen3-8B and $24.7\%$ for GPT-OSS-20B; the
exception is Granite-3.3-8B, whose majority deciders exclude more often at this
size, $22.2\%$ against $9.5\%$). At the most lopsided size the ordering does not hold
in any event: majority deciders exclude the
out-group more often than minority deciders in four of the six models
(R1-Distill-Llama-8B $2.7\%$ vs.\ $2.2\%$, Phi-4-reasoning $9.6\%$ vs.\ $5.6\%$,
R1-Distill-Qwen-14B $21.4\%$ vs.\ $10.1\%$, and Granite-3.3-8B $44.0\%$ vs.\
$8.3\%$), with Qwen3-8B and GPT-OSS-20B running the other way. Granite-3.3-8B is
the extreme case of that pattern rather than the only instance of it: its
$44.0\%$ is the single highest rate anywhere in the table, more than five times
its own minority rate at that size, and it declines monotonically to $7.8\%$ at
the equal split. Because a fully excluding allocation is far easier to produce
when the out-group is small, we read this categorical measure as descriptive
only and rest the size comparisons on the excess-share measure, which is
constructed to be size-robust. None of this is the
sign-reversal seen for R1-Distill-Llama-8B (Appendix~\ref{sm:reversal}): Granite's
mean excess share for majority deciders at this size stays positive ($+0.077$;
Table~\ref{tab:share}), so a substantial share of its majority deciders give the
entire pool to their own group while the rest split closer to evenly, averaging to
a modest net in-group excess rather than reversing it. In the group-blind control
the behavior is rare across every model and decider group ($\leq2.1\%$) -- a
categorical version of the same validation the graded measure provides. As with
the excess-share measure (Appendix~\ref{sm:stats}), the rate is not monotone in
group size for most model/decider-group cells (8 of 12), and for minority
deciders it peaks at the intermediate minority in five of the six models; we
therefore report it as a
descriptive robustness check rather than a graded dependence on size.

\begin{table*}[t]
 \caption{Rate of allocations excluding all out-group targets (zero points to every
 out-group member), by minority size, labels visible; the right column is the
 group-blind control (pooled over size). Rates are percentages of parsed
 allocations.}
 \centering
 \small
 \begin{tabular}{ll rrr r}
   \toprule
   & & \multicolumn{3}{c}{Labels visible, minority size} & Group-blind \\
   \cmidrule(lr){3-5}\cmidrule(lr){6-6}
   Model & Decider & $3$ & $5$ & $10$ & \\
   \midrule
R1-Distill-Llama-8B & minority & 2.2\% & 7.0\% & 1.7\% & 0.0\% \\
 & majority & 2.7\% & 2.9\% & 1.5\% & 1.0\% \\
\midrule
Phi-4-reasoning (14B) & minority & 5.6\% & 8.1\% & 8.4\% & 0.0\% \\
 & majority & 9.6\% & 6.3\% & 7.4\% & 0.0\% \\
\midrule
Qwen3-8B & minority & 10.0\% & 30.4\% & 4.0\% & 0.0\% \\
 & majority & 3.8\% & 1.6\% & 4.2\% & 0.0\% \\
\midrule
R1-Distill-Qwen-14B & minority & 10.1\% & 18.8\% & 3.5\% & 0.0\% \\
 & majority & 21.4\% & 9.7\% & 4.0\% & 1.2\% \\
\midrule
Granite-3.3-8B & minority & 8.3\% & 9.5\% & 9.3\% & 0.0\% \\
 & majority & 44.0\% & 22.2\% & 7.8\% & 0.2\% \\
\midrule
GPT-OSS-20B & minority & 12.0\% & 24.7\% & 4.4\% & 0.7\% \\
 & majority & 9.1\% & 6.9\% & 4.7\% & 2.1\% \\
\midrule
   \bottomrule
 \end{tabular}
 \label{tab:exclusion}
\end{table*}

\subsection{Multiple-comparison correction and effect sizes}
\label{sm:fdr}
The per-cell seed-level tests form a family of $72$ one-sample Wilcoxon tests ($6$
models $\times$ $2$ decider groups $\times$ $3$ sizes $\times$ $2$ label
conditions), taken here one-sided against the directional hypothesis the paper
advances, that allocation favors the in-group. Table~\ref{tab:percell} reports the
same cells two-sided, so the two are not interchangeable: a cell displaced strongly
in the \emph{opposite} direction is highly significant there and counts as not
surviving here. Exactly one cell is of that kind -- R1-Distill-Llama-8B's
majority-decider cell at minority size $3$ (Appendix~\ref{sm:reversal}) -- and we
flag it rather than let the count imply a null. Applying a Benjamini--Hochberg
correction across all $72$, $33$ of the $36$ visible cells remain significant at
$q<0.05$ (largest surviving $q=0.04$), while only $1$ of the $36$ group-blind cells
does -- the false discovery rate expected by chance. Table~\ref{tab:fdr} breaks this down by decider
group and adds a standardized effect size, the matched-pairs rank-biserial
correlation. Every minority-decider block survives correction in full ($3/3$ cells)
with near-maximal effect sizes ($r$ up to $+1.00$); the three visible cells that do
not survive are all majority-decider cells, all in two models
(R1-Distill-Llama-8B and R1-Distill-Qwen-14B), and the single
group-blind cell that crosses is likewise a majority cell -- consistent with the
effect being concentrated in the minority.

\begin{table*}[t]
 \caption{Multiple-comparison correction and effect sizes. ``Visible cells $q<0.05$''
 is the number of the three per-size seed-level Wilcoxon tests (per model and decider
 group), taken one-sided in the in-group-favoring direction, that survive
 Benjamini--Hochberg correction across the full family of $72$ tests; ``rank-biserial $r$'' is the range of the matched-pairs effect size over
 those three cells; the last column is the same count for the group-blind arm.}
 \centering
 \small
 \begin{tabular}{ll c c c}
   \toprule
   & & Visible cells & Rank-biserial $r$ & Group-blind cells \\
   Model & Decider & $q<0.05$ & (range) & $q<0.05$ \\
   \midrule
R1-Distill-Llama-8B & minority & 3/3 & $+0.97$ to $+1.00$ & 0/3 \\
 & majority & 1/3 & $-1.00$ to $+1.00$ & 1/3 \\
\midrule
Phi-4-reasoning (14B) & minority & 3/3 & $+1.00$ to $+1.00$ & 0/3 \\
 & majority & 3/3 & $+0.95$ to $+1.00$ & 0/3 \\
\midrule
Qwen3-8B & minority & 3/3 & $+1.00$ to $+1.00$ & 0/3 \\
 & majority & 3/3 & $+0.49$ to $+1.00$ & 0/3 \\
\midrule
R1-Distill-Qwen-14B & minority & 3/3 & $+1.00$ to $+1.00$ & 0/3 \\
 & majority & 2/3 & $+0.23$ to $+1.00$ & 0/3 \\
\midrule
Granite-3.3-8B & minority & 3/3 & $+1.00$ to $+1.00$ & 0/3 \\
 & majority & 3/3 & $+1.00$ to $+1.00$ & 0/3 \\
\midrule
GPT-OSS-20B & minority & 3/3 & $+1.00$ to $+1.00$ & 0/3 \\
 & majority & 3/3 & $+0.44$ to $+1.00$ & 0/3 \\
\midrule
   \bottomrule
 \end{tabular}
 \label{tab:fdr}
\end{table*}

\subsection{Seed-level cluster bootstrap}
\label{sm:bootstrap}
The dyad-level GEE (Table~\ref{tab:gee}) clusters its sandwich standard error on the
$30$ seeds of a model$\times$condition fit -- fewer than the $\sim$40--50 clusters
conventionally recommended for the sandwich estimator's asymptotic justification, so
an analytic SE from that few clusters could in principle be anti-conservative. Under
an identity link with an independence working correlation and no covariate besides
\texttt{same\_group}, the GEE point estimate is the same in-group-minus-out-group
difference in means that a nonparametric bootstrap can reproduce without relying on
sandwich asymptotics at all. We therefore resampled the $30$ seeds with replacement
($2{,}000$ resamples), recomputed the same\_group difference-in-means (and the
minority$-$majority double difference) on each resample, and compared the resulting
bootstrap standard error to the analytic sandwich SE. Table~\ref{tab:bootstrap} shows
they agree to within $0.003$ points per target for every model and both quantities, and
every bootstrap 95\% percentile interval excludes zero by a wide margin. The small
cluster count is therefore not an alternative explanation for the reported
significance; we report it here as a direct check rather than relying only on the
sandwich estimator's asymptotics, alongside the seed-level Wilcoxon tests already
described in Appendix~\ref{sm:stats}, which do not depend on cluster asymptotics at
all.

\begin{table*}[t]
 \caption{Seed-level cluster bootstrap corroborates the analytic sandwich SE.
 ``Analytic SE'' is the dyad-level GEE robust SE from Table~\ref{tab:gee} (seed
 clustered, $n=30$ clusters); ``Bootstrap SE'' is the standard deviation of the same
 statistic recomputed over $2{,}000$ seed-level bootstrap resamples; ``Bootstrap 95\%
 CI'' is the corresponding percentile interval. Left block: the same\_group
 coefficient (in-group favoritism, points per target). Right block: the
 same\_group$\times$decider-group interaction (minority$-$majority gap).}
 \centering
 \small
 \resizebox{\textwidth}{!}{%
 \begin{tabular}{l rrrl rrrl}
   \toprule
   & \multicolumn{4}{c}{In-group favoritism} & \multicolumn{4}{c}{Minority$-$majority gap} \\
   \cmidrule(lr){2-5}\cmidrule(lr){6-9}
   Model & $\beta$ & Analytic SE & Bootstrap SE & Bootstrap 95\% CI
         & $\beta$ & Analytic SE & Bootstrap SE & Bootstrap 95\% CI \\
   \midrule
R1-Distill-Llama-8B & +0.839 & 0.079 & 0.079 & [+0.68, +0.99] & +2.383 & 0.171 & 0.174 & [+2.06, +2.73] \\
Phi-4-reasoning (14B) & +2.088 & 0.063 & 0.064 & [+1.96, +2.21] & +2.473 & 0.158 & 0.161 & [+2.16, +2.79] \\
Qwen3-8B & +2.314 & 0.066 & 0.064 & [+2.19, +2.44] & +4.083 & 0.183 & 0.182 & [+3.73, +4.44] \\
R1-Distill-Qwen-14B & +4.064 & 0.066 & 0.065 & [+3.94, +4.20] & +6.841 & 0.206 & 0.209 & [+6.43, +7.23] \\
Granite-3.3-8B & +3.031 & 0.049 & 0.049 & [+2.93, +3.13] & +2.820 & 0.164 & 0.166 & [+2.49, +3.14] \\
GPT-OSS-20B & +2.222 & 0.068 & 0.067 & [+2.09, +2.35] & +4.182 & 0.198 & 0.200 & [+3.80, +4.58] \\
   \bottomrule
 \end{tabular}%
 }
 \label{tab:bootstrap}
\end{table*}

\subsection{Paraphrase-instrument robustness}
\label{sm:paraphrase}
The allocation instrument's prompt is one specific wording of the Tajfel task; to
check that the reported behavior is not an artifact of that particular phrasing, we
reran a reduced design under a semantically equivalent \emph{paraphrase} of both the
system and user prompts (\texttt{scripts/orchestrate\_paraphrase.py}; comparison by
\texttt{analysis/make\_paraphrase\_check.py}). The paraphrase keeps every structural
element of the task identical -- the point pool, the JSON output contract, and the
roster/label formatting (Appendix~\ref{sm:prompts}) -- and differs in surface
wording, e.g.\ the system prompt opens ``You take part in a shared social setting
alongside other participants\ldots'' in place of ``You are a participant in a social
environment with other participants\ldots'', and the user message opens ``Your
participant id is \textless decider\_id\textgreater'' in place of ``You are
\textless decider\_id\textgreater''. Every sentence of the original has a
counterpart, including the inert ``You have a personality.'', rendered ``You
have a personality of your own.'' An earlier version of this check used a
paraphrase that omitted that sentence, so it confounded a rewording with a
deletion; it is superseded by the collection reported here, and the one result
it produced that this one does not is noted below.

The reduced design covers the sharpest-asymmetry and equal-split visible
conditions (minority sizes~$3$ and~$10$) plus the hidden-label floor (size~$3$),
for all six models, at the first $15$ of the original study's $30$ seeds
($9{,}000$ trials). A society's composition is drawn from
the integer seed alone, so seeds $0$--$14$ reproduce the original study's group
memberships and label pairs exactly (verified trial by trial: $100\%$ agreement
on both). What does \emph{not} carry over is the order in which each decider's
roster is presented, which was drawn from a process-salted hash and so differs
between the two runs on every trial. The pairing is therefore matched on society
composition but not on presentation order, which is the weaker of the two and
sufficient here: roster order is randomized by design, and
Appendix~\ref{sm:position} shows in-group and out-group targets occupy mean
positions within $0.07$ of each other, so the primacy effect balances across the
group boundary under either draw. On that pairing
(Table~\ref{tab:paraphrase}) we compare, for each seed, that seed's
original-instrument favoritism to the same seed's paraphrase-instrument
favoritism (Wilcoxon signed-rank), and separately refit the dyad-level GEE on each
instrument's trials.

The qualitative pattern reported in the main text survives the rewording. Of the
$36$ visible-condition cells ($6$ models~$\times$~$2$ sizes~$\times$~$3$ decider
groups), $32$ retain the same sign (positive favoritism under both instruments).
The four that do not are all at minority size~$3$: two majority-decider cells
whose original value lies within $0.4$ points per target of zero (GPT-OSS-20B,
R1-Distill-Qwen-14B), the R1-Distill-Llama-8B majority cell that is the reversal
of Appendix~\ref{sm:reversal}, and that model's society-level cell, which the
reversal drives. The dyad-level GEE same\_group coefficient -- the effect-size
metric reported in Table~\ref{tab:gee} -- is positive under the paraphrase in
all twelve visible model$\times$size cells. The two instruments disagree in sign
in exactly one of those twelve: R1-Distill-Llama-8B at minority size~$3$, where
the original gives $\beta=-0.33$ (SE~$0.19$) on this 15-seed subsample -- the
weakest and noisiest visible-condition cell under either instrument -- against
$\beta=+0.87$ (SE~$0.19$) under the paraphrase. That is a magnitude shift in the
weakest cell rather than the reversal of an otherwise robust effect. The size
dependence also survives: at minority size~$3$, minority deciders allocate more
to their in-group than majority deciders under the paraphrase in every model
($+4.5$ to $+14.6$ against $-0.04$ to $+2.5$ points per target). Exact
magnitudes are wording-sensitive in both directions. Paraphrase favoritism is
smaller than the original in four models, by up to a factor of three (GEE
$\beta$ at size~$10$: GPT-OSS-20B $2.85\rightarrow0.92$, R1-Distill-Qwen-14B
$6.99\rightarrow4.93$, Granite-3.3-8B $3.26\rightarrow2.49$, Qwen3-8B
$4.05\rightarrow3.43$), and larger in two (R1-Distill-Llama-8B at size~$10$,
$1.59\rightarrow2.95$; Phi-4-reasoning at size~$3$, $1.26\rightarrow1.70$);
$24$ of the $36$ visible cells differ significantly between instruments
($p<0.05$, seed-level Wilcoxon). With only $15$ matched seeds this is the
sensitivity expected of a single fixed prompt template, of a piece with the
wide spread language models show across semantically equivalent prompt formats
and example orderings \citep{sclar2024quantifying,lu2022fantastically}, rather than evidence against
the underlying effect, which is directionally and statistically robust to the
rewording.

The group-blind floor is unaffected by the rewording in every model. Under the
paraphrase the society-level floor sits between $+0.01$ and $+0.29$ points per
target and the dyad-level GEE floor between $+0.01$ and $+0.19$, against visible
GEE effects of $0.44$ to $4.93$ from the same instrument, and none of the $18$
hidden-condition cells differs significantly between instruments (smallest
$p=0.15$). The largest movement is R1-Distill-Llama-8B's, from $+0.01$ to
$+0.29$ at the society level (GEE $+0.19$, SE~$0.11$), which is within the
noise of a 15-seed cell and about a third of that model's smallest
visible GEE effect under the paraphrase ($+0.87$). Under the earlier,
confounded version of this check the same model's floor had moved to $+0.82$
($p=0.012$ against the original); with the inert sentence restored it does
not, so that movement was a response to the deletion rather than to the
rewording. What the check establishes is that the visible effect, its size
dependence and the floor it is read against all survive a rewording of both
prompts in all six models, with magnitudes that are wording-sensitive by up to
a factor of three.

\begin{table*}[t]
 \caption{Paraphrase-instrument robustness: seed-matched comparison ($n=15$ matched
 seeds per cell) of in-group favoritism (points per target, mean per seed) under the
 original instrument reported in the main text versus a semantically equivalent
 paraphrase. ``Decider'' is society (all deciders pooled), or the minority-/
 majority-decider subset. $p$ is a two-sided Wilcoxon signed-rank test on the
 within-seed original$-$paraphrase difference. \texttt{visible\_minority03} and
 \texttt{visible\_minority10} are the sharpest-asymmetry and equal-split visible
 conditions; \texttt{hidden\_minority03} is the group-blind floor. The corresponding
 dyad-level GEE comparison is discussed in Appendix~\ref{sm:paraphrase}.}
 \centering
 \footnotesize
 \setlength{\tabcolsep}{4pt}\renewcommand{\arraystretch}{0.92}
 \begin{tabular}{l l l r rr rl}
   \toprule
   Model & Condition & Decider & $n$ & Orig. & Paraphrase & $\Delta$ & $p$ \\
   \midrule
R1-Distill-Llama-8B & \texttt{hidden\_minority03} & society & 15 & +0.01 & +0.29 & -0.28 & 0.1876 \\
R1-Distill-Llama-8B & \texttt{hidden\_minority03} & minority & 15 & +0.04 & +0.27 & -0.24 & 0.9547 \\
R1-Distill-Llama-8B & \texttt{hidden\_minority03} & majority & 15 & +0.01 & +0.30 & -0.29 & 0.3894 \\
R1-Distill-Llama-8B & \texttt{visible\_minority03} & society & 15 & -0.18 & +1.72 & -1.90 & 0.0034 \\
R1-Distill-Llama-8B & \texttt{visible\_minority03} & minority & 15 & +7.02 & +10.04 & -3.01 & 0.3303 \\
R1-Distill-Llama-8B & \texttt{visible\_minority03} & majority & 15 & -1.45 & +0.25 & -1.70 & $<10^{-4}$ \\
R1-Distill-Llama-8B & \texttt{visible\_minority10} & society & 15 & +1.59 & +2.95 & -1.37 & 0.0003 \\
R1-Distill-Llama-8B & \texttt{visible\_minority10} & minority & 15 & +1.44 & +3.21 & -1.77 & 0.0020 \\
R1-Distill-Llama-8B & \texttt{visible\_minority10} & majority & 15 & +1.74 & +2.70 & -0.96 & 0.0084 \\
Phi-4-reasoning (14B) & \texttt{hidden\_minority03} & society & 15 & +0.02 & +0.02 & +0.01 & 0.6788 \\
Phi-4-reasoning (14B) & \texttt{hidden\_minority03} & minority & 15 & +0.10 & +0.08 & +0.02 & 0.7775 \\
Phi-4-reasoning (14B) & \texttt{hidden\_minority03} & majority & 15 & +0.01 & +0.01 & +0.00 & 0.5245 \\
Phi-4-reasoning (14B) & \texttt{visible\_minority03} & society & 15 & +2.36 & +2.98 & -0.61 & 0.2078 \\
Phi-4-reasoning (14B) & \texttt{visible\_minority03} & minority & 15 & +11.48 & +10.60 & +0.88 & 0.8469 \\
Phi-4-reasoning (14B) & \texttt{visible\_minority03} & majority & 15 & +0.75 & +1.63 & -0.88 & 0.0009 \\
Phi-4-reasoning (14B) & \texttt{visible\_minority10} & society & 15 & +3.27 & +3.31 & -0.05 & 0.5614 \\
Phi-4-reasoning (14B) & \texttt{visible\_minority10} & minority & 15 & +3.26 & +3.28 & -0.02 & 0.9780 \\
Phi-4-reasoning (14B) & \texttt{visible\_minority10} & majority & 15 & +3.27 & +3.34 & -0.07 & 0.8904 \\
Qwen3-8B & \texttt{hidden\_minority03} & society & 15 & +0.00 & +0.01 & -0.01 & 0.9780 \\
Qwen3-8B & \texttt{hidden\_minority03} & minority & 15 & -0.07 & -0.03 & -0.03 & 0.7004 \\
Qwen3-8B & \texttt{hidden\_minority03} & majority & 15 & +0.02 & +0.02 & -0.00 & 0.8647 \\
Qwen3-8B & \texttt{visible\_minority03} & society & 15 & +1.60 & +0.90 & +0.69 & 0.0125 \\
Qwen3-8B & \texttt{visible\_minority03} & minority & 15 & +9.81 & +4.49 & +5.32 & 0.0054 \\
Qwen3-8B & \texttt{visible\_minority03} & majority & 15 & +0.15 & +0.27 & -0.12 & 0.3894 \\
Qwen3-8B & \texttt{visible\_minority10} & society & 15 & +4.05 & +3.43 & +0.62 & 0.0003 \\
Qwen3-8B & \texttt{visible\_minority10} & minority & 15 & +3.91 & +3.83 & +0.08 & 0.9780 \\
Qwen3-8B & \texttt{visible\_minority10} & majority & 15 & +4.20 & +3.03 & +1.17 & 0.0006 \\
R1-Distill-Qwen-14B & \texttt{hidden\_minority03} & society & 15 & +0.03 & +0.06 & -0.03 & 0.8040 \\
R1-Distill-Qwen-14B & \texttt{hidden\_minority03} & minority & 15 & -0.22 & -0.17 & -0.05 & 0.7532 \\
R1-Distill-Qwen-14B & \texttt{hidden\_minority03} & majority & 15 & +0.07 & +0.10 & -0.02 & 0.5995 \\
R1-Distill-Qwen-14B & \texttt{visible\_minority03} & society & 15 & +3.84 & +2.16 & +1.67 & 0.0084 \\
R1-Distill-Qwen-14B & \texttt{visible\_minority03} & minority & 15 & +23.55 & +14.63 & +8.92 & 0.0009 \\
R1-Distill-Qwen-14B & \texttt{visible\_minority03} & majority & 15 & +0.36 & -0.04 & +0.40 & 0.3591 \\
R1-Distill-Qwen-14B & \texttt{visible\_minority10} & society & 15 & +6.99 & +4.93 & +2.06 & $<10^{-4}$ \\
R1-Distill-Qwen-14B & \texttt{visible\_minority10} & minority & 15 & +6.98 & +4.68 & +2.29 & $<10^{-4}$ \\
R1-Distill-Qwen-14B & \texttt{visible\_minority10} & majority & 15 & +7.00 & +5.18 & +1.82 & 0.0002 \\
Granite-3.3-8B & \texttt{hidden\_minority03} & society & 15 & +0.08 & +0.13 & -0.05 & 0.8040 \\
Granite-3.3-8B & \texttt{hidden\_minority03} & minority & 15 & +0.09 & +0.20 & -0.10 & 0.8904 \\
Granite-3.3-8B & \texttt{hidden\_minority03} & majority & 15 & +0.08 & +0.12 & -0.04 & 0.6788 \\
Granite-3.3-8B & \texttt{visible\_minority03} & society & 15 & +5.65 & +3.41 & +2.23 & $<10^{-4}$ \\
Granite-3.3-8B & \texttt{visible\_minority03} & minority & 15 & +20.62 & +8.79 & +11.84 & 0.0001 \\
Granite-3.3-8B & \texttt{visible\_minority03} & majority & 15 & +3.00 & +2.46 & +0.54 & 0.3894 \\
Granite-3.3-8B & \texttt{visible\_minority10} & society & 15 & +3.26 & +2.49 & +0.76 & 0.0020 \\
Granite-3.3-8B & \texttt{visible\_minority10} & minority & 15 & +3.53 & +2.32 & +1.22 & 0.0009 \\
Granite-3.3-8B & \texttt{visible\_minority10} & majority & 15 & +2.98 & +2.67 & +0.31 & 0.3894 \\
GPT-OSS-20B & \texttt{hidden\_minority03} & society & 15 & -0.09 & +0.04 & -0.12 & 0.2769 \\
GPT-OSS-20B & \texttt{hidden\_minority03} & minority & 15 & -0.31 & +0.20 & -0.51 & 0.1547 \\
GPT-OSS-20B & \texttt{hidden\_minority03} & majority & 15 & -0.05 & +0.01 & -0.05 & 0.7615 \\
GPT-OSS-20B & \texttt{visible\_minority03} & society & 15 & +1.92 & +0.88 & +1.04 & 0.0353 \\
GPT-OSS-20B & \texttt{visible\_minority03} & minority & 15 & +12.81 & +5.38 & +7.43 & 0.0006 \\
GPT-OSS-20B & \texttt{visible\_minority03} & majority & 15 & -0.01 & +0.09 & -0.09 & 1.0000 \\
GPT-OSS-20B & \texttt{visible\_minority10} & society & 15 & +2.85 & +0.92 & +1.93 & $<10^{-4}$ \\
GPT-OSS-20B & \texttt{visible\_minority10} & minority & 15 & +2.88 & +0.98 & +1.90 & $<10^{-4}$ \\
GPT-OSS-20B & \texttt{visible\_minority10} & majority & 15 & +2.81 & +0.86 & +1.95 & 0.0003 \\
   \bottomrule
 \end{tabular}
 \label{tab:paraphrase}
\end{table*}

\subsection{Favoritism against the range each cell can reach}
\label{sm:normshare}
Excess share is $0$ for a per-capita-even allocation at
any group size, which is what it was introduced for, but it is bounded above
in a way that depends on group size. A decider cannot raise its in-group's
share of the pool above $1$, so its excess share cannot exceed
\[
  C(n_{\text{in}}) = 1 - n_{\text{in}}/19,
\]
attained by giving that group the whole pool. The ceiling is $17/19=0.895$ for a minority decider at minority
size $3$ and $3/19=0.158$ for a majority decider in the same society, so those
two cells are not on a common scale and part of any raw difference between
them is fixed by the head counts alone. Table~\ref{tab:normshare} therefore
reports each cell both raw and divided by its ceiling, the latter being the
fraction of the attainable range the cell uses, with a seed-level bootstrap
interval ($10{,}000$ resamples of the $30$ societies).

Three readings follow, and Section~\ref{sec:asymmetry} rests on them. At the
equal $10{:}10$ split, where both decider groups face $n_{\text{in}}=9$ and
the rescaling is the same for both, the two sides are within $0.04$ of each
other in every model and sit between $0.15$ and $0.65$ of the range. Moving to
the unequal splits does not, in general, lift the minority side above its own
parity value: in R1-Distill-Llama-8B, Phi-4-reasoning, Qwen3-8B and
R1-Distill-Qwen-14B the smallest minority uses a smaller fraction of its range
than a decider at parity does, and only in Granite-3.3-8B does the normalized
minority cell rise steadily as the minority shrinks. The majority side is where the size dependence lives: in five of
the six models it falls monotonically as the decider's own group grows from
$9$ to $14$ to $16$ targets, from $0.17$--$0.62$ at parity to $-0.23$--$+0.13$
at the most lopsided split. Granite-3.3-8B is the sole exception and moves the
other way ($0.276$, $0.350$, $0.488$), which is the same fact as its being the
one model not separated from size-independent matching
(Appendix~\ref{sm:matching}).

Two limits on this table. Dividing by the ceiling is a descriptive rescaling
that puts the two cells on a comparable $[0,1]$ scale; it is not a model of
how the allocations are produced, and it is not the only defensible rescaling.
The inferential version of the same correction is Appendix~\ref{sm:matching},
which derives the lowest majority value compatible with any rule that cannot
see the head counts and tests each cell against it; the two are algebraically
the same statement, since that appendix's shortfall equals
$C(n_{\text{in}}^{\text{maj}})$ times the normalized gap
$E_{\text{min}}/C_{\text{min}} - E_{\text{maj}}/C_{\text{maj}}$ reported here.
A cell whose normalized majority value falls below its normalized minority
value is exactly a cell with a positive shortfall there, and the interval that
decides significance is the one in Table~\ref{tab:matching}.

\begin{table*}[t]
 \caption{In-group excess share against the range each cell can reach, labels
 visible. $n_{\text{in}}$ is the number of same-tag targets a decider of that
 group faces at that minority size and $C=1-n_{\text{in}}/19$ is the largest
 excess share it could attain, reached by giving its own group the entire pool.
 ``Excess share'' is the observed value (Table~\ref{tab:share}); ``of range'' is
 that value divided by $C$, with a $95\%$ percentile interval from resampling
 the $30$ societies of the cell ($10{,}000$ resamples). At the equal $10{:}10$
 split both decider groups face $n_{\text{in}}=9$, so the rescaling is common to
 them and the raw and normalized comparisons agree. Generated by
 \texttt{analysis/make\_norm\_share.py}.}
 \centering
 \small
 \begin{tabular}{ll l rr rr l}
   \toprule
   Model & Size & Decider & $n_{\text{in}}$ & $C$ & Excess share & of range & $95\%$ CI \\
   \midrule
R1-Distill-Llama-8B & 3 & minority & 2 & 0.895 & +0.113 & +0.126 & [+0.085, +0.170] \\
 &  & majority & 16 & 0.158 & -0.037 & -0.231 & [-0.291, -0.175] \\
 & 5 & minority & 4 & 0.789 & +0.166 & +0.210 & [+0.179, +0.242] \\
 &  & majority & 14 & 0.263 & +0.003 & +0.011 & [-0.020, +0.041] \\
 & 10 & minority & 9 & 0.526 & +0.081 & +0.153 & [+0.114, +0.193] \\
 &  & majority & 9 & 0.526 & +0.089 & +0.169 & [+0.146, +0.192] \\
\midrule
Phi-4-reasoning (14B) & 3 & minority & 2 & 0.895 & +0.186 & +0.208 & [+0.165, +0.252] \\
 &  & majority & 16 & 0.158 & +0.021 & +0.130 & [+0.097, +0.163] \\
 & 5 & minority & 4 & 0.789 & +0.219 & +0.278 & [+0.237, +0.323] \\
 &  & majority & 14 & 0.263 & +0.055 & +0.210 & [+0.189, +0.232] \\
 & 10 & minority & 9 & 0.526 & +0.146 & +0.278 & [+0.252, +0.304] \\
 &  & majority & 9 & 0.526 & +0.140 & +0.266 & [+0.243, +0.292] \\
\midrule
Qwen3-8B & 3 & minority & 2 & 0.895 & +0.161 & +0.180 & [+0.131, +0.235] \\
 &  & majority & 16 & 0.158 & +0.005 & +0.032 & [+0.009, +0.060] \\
 & 5 & minority & 4 & 0.789 & +0.364 & +0.461 & [+0.418, +0.504] \\
 &  & majority & 14 & 0.263 & +0.011 & +0.041 & [+0.021, +0.061] \\
 & 10 & minority & 9 & 0.526 & +0.185 & +0.352 & [+0.320, +0.385] \\
 &  & majority & 9 & 0.526 & +0.191 & +0.363 & [+0.341, +0.385] \\
\midrule
R1-Distill-Qwen-14B & 3 & minority & 2 & 0.895 & +0.416 & +0.465 & [+0.430, +0.499] \\
 &  & majority & 16 & 0.158 & +0.007 & +0.042 & [-0.047, +0.127] \\
 & 5 & minority & 4 & 0.789 & +0.493 & +0.624 & [+0.590, +0.658] \\
 &  & majority & 14 & 0.263 & +0.041 & +0.155 & [+0.111, +0.198] \\
 & 10 & minority & 9 & 0.526 & +0.341 & +0.648 & [+0.620, +0.675] \\
 &  & majority & 9 & 0.526 & +0.324 & +0.616 & [+0.589, +0.641] \\
\midrule
Granite-3.3-8B & 3 & minority & 2 & 0.895 & +0.379 & +0.424 & [+0.381, +0.465] \\
 &  & majority & 16 & 0.158 & +0.077 & +0.488 & [+0.435, +0.540] \\
 & 5 & minority & 4 & 0.789 & +0.267 & +0.338 & [+0.306, +0.371] \\
 &  & majority & 14 & 0.263 & +0.092 & +0.350 & [+0.319, +0.380] \\
 & 10 & minority & 9 & 0.526 & +0.165 & +0.314 & [+0.285, +0.341] \\
 &  & majority & 9 & 0.526 & +0.145 & +0.276 & [+0.247, +0.305] \\
\midrule
GPT-OSS-20B & 3 & minority & 2 & 0.895 & +0.266 & +0.298 & [+0.248, +0.352] \\
 &  & majority & 16 & 0.158 & +0.005 & +0.032 & [-0.009, +0.070] \\
 & 5 & minority & 4 & 0.789 & +0.375 & +0.475 & [+0.433, +0.516] \\
 &  & majority & 14 & 0.263 & +0.032 & +0.120 & [+0.095, +0.145] \\
 & 10 & minority & 9 & 0.526 & +0.145 & +0.276 & [+0.247, +0.304] \\
 &  & majority & 9 & 0.526 & +0.138 & +0.262 & [+0.239, +0.284] \\
\midrule
   \bottomrule
 \end{tabular}
 \label{tab:normshare}
\end{table*}

\subsection{Ruling out size-independent token matching}
\label{sm:matching}
Section~\ref{sec:origin} sets the main effect against the reductive reading
that the models match a string rather than condition on group structure: give a
target more weight when its tag equals the decider's own. That heuristic is not
hypothetical -- a forced-choice audit finds models up-weighting a candidate that
shares an incidental token with their context \citep{anon_companion} -- so it has
to be excluded rather than set aside. The exclusion below is self-contained: it
is derived and evaluated on this paper's own allocations, and nothing in it
depends on that companion result.

Write it down. A target receives weight $r$ if its tag matches the decider's and
$1$ if it does not, and the pool is divided in proportion. Only the ratio
survives that normalization, so this is the general form: any rule that scores a
target on match versus no-match alone reduces to some $r$. Its defining property
is that $r$ cannot depend on how many agents carry each tag, because a string
match does not see the head counts. For a decider with $n_{\text{in}}$ same-tag
and $n_{\text{out}}$ other-tag targets the in-group's excess share is then
\[
  E(n_{\text{in}}, n_{\text{out}}, r)
  = \frac{n_{\text{in}} r}{n_{\text{in}} r + n_{\text{out}}}
  - \frac{n_{\text{in}}}{n_{\text{in}} + n_{\text{out}}},
\]
zero at $r=1$ and increasing in $r$.

This does \emph{not} predict a constant effect. At a fixed $r$ the excess share
still depends on the head counts, and it is larger for a decider in the smaller
group, so a matching rule reproduces the qualitative shape of
Fig.~\ref{fig:dose} on its own. The measure is size-robust in the sense that it
is zero for a per-capita-even allocation at any group size, which is what it was
introduced for; it is not free of group size everywhere in its range, and the
minority-above-majority ordering by itself is therefore not evidence against
matching. The test has to be quantitative.

It is available because minority and majority deciders in the same society face
different $(n_{\text{in}}, n_{\text{out}})$ while, under this account, drawing
$r$ from one distribution. As $r$ sweeps $1$ to $\infty$ the pair
$(E_{\text{min}}, E_{\text{maj}})$ traces a curve from $(0,0)$ to
$(1-n_{\text{in}}^{\text{min}}/19,\ 1-n_{\text{in}}^{\text{maj}}/19)$, and that
curve is concave, the majority coordinate saturating first because a majority
decider's in-group already holds most of the head count. A mixture over $r$ --
trials varying in how hard they match -- gives an expectation of points on the
curve and so lies in its convex hull, whose lower boundary, for a concave curve,
is the chord between the endpoints. Hence for any $r$ and any mixture over $r$,
\[
  E_{\text{maj}} \;\ge\; \frac{1-n_{\text{in}}^{\text{maj}}/19}
                              {1-n_{\text{in}}^{\text{min}}/19}\; E_{\text{min}},
\]
which is $0.176\,E_{\text{min}}$ at minority size $3$ and $0.333\,E_{\text{min}}$
at size $5$. A majority cell below its floor cannot be produced by any size-independent
matching rule, however strong and however heterogeneous across trials.
Equivalently, the bound says that the two decider groups of a society use the
same fraction of the excess share each could attain: the shortfall below it is
$C(n_{\text{in}}^{\text{maj}})$ times the ceiling-normalized gap of
Appendix~\ref{sm:normshare}, which is the descriptive form of the same
comparison.

Table~\ref{tab:matching} evaluates it. Ten of the twelve unequal-split cells
fall below the floor, several by a wide margin, and in all ten a $95\%$
percentile interval from resampling the cell's $30$ societies excludes zero --
including Phi-4-reasoning's two, the narrowest shortfalls of the ten
($+0.012$, CI $[+0.003,+0.022]$; $+0.018$, CI $[+0.006,+0.030]$). The two cells
that do not fall below are Granite-3.3-8B's, which clear the floor by $0.010$
and $0.003$ with intervals straddling zero; that model's behavior at these
sizes is not separated from what size-independent matching can produce, and we
do not claim otherwise for it. Granite-3.3-8B is also the model whose majority deciders
exclude the out-group entirely most often ($44.0\%$ at minority size $3$;
Appendix~\ref{sm:exclusion}), which is the same behavior seen from a different
angle: its majority deciders favor their own group about as hard as its minority
deciders do, which is what a size-blind rule predicts and what the other five
models do not show.

Two limits on the argument. The bound assumes $r\ge1$, matching that favors the
in-group, which is the direction the heuristic is claimed in; admitting trials
with $r<1$ enlarges the hull and weakens it, and R1-Distill-Llama-8B's negative
majority cell at minority size~3 is the one place that plainly matters
(Appendix~\ref{sm:reversal}). And ruling out this account does not identify the
one that replaces it: it establishes that the effective strength of the
favoritism varies with the composition of the society, not what the models
compute to make it do so.

\begin{table*}[t]
 \caption{Size-independent token matching cannot produce ten of the twelve
 unequal-split cells. ``Minority excess'' is the observed in-group excess share
 for minority deciders (Table~\ref{tab:share}); ``majority floor'' is the least
 majority excess share compatible with \emph{any} size-independent matching rule
 given that minority value, and any mixture of such rules across trials;
 ``majority excess'' is what was observed. ``Shortfall'' is the floor minus the
 observed value, with a $95\%$ percentile interval from resampling the $30$
 societies of the cell with replacement ($10{,}000$ resamples; minority and
 majority excess share are recomputed together on each resample, since they come
 from the same societies). A cell falsifies the account when its shortfall is
 positive and that interval excludes zero (in bold); an interval straddling
 zero means the cell is not separated from the floor, not that the account is
 confirmed. The equal $10{:}10$ split is omitted: both decider groups face the
 same head counts there, so the account predicts they agree, and they do.}
 \centering
 \small
 \begin{tabular}{ll rrr rl}
   \toprule
   Model & Size & Minority excess & Majority floor & Majority excess & Shortfall & $95\%$ CI \\
   \midrule
R1-Distill-Llama-8B & 3 & +0.113 & +0.020 & -0.037 & \textbf{+0.056} & [+0.045, +0.069] \\
 & 5 & +0.166 & +0.055 & +0.003 & \textbf{+0.052} & [+0.041, +0.064] \\
\midrule
Phi-4-reasoning (14B) & 3 & +0.186 & +0.033 & +0.021 & \textbf{+0.012} & [+0.003, +0.022] \\
 & 5 & +0.219 & +0.073 & +0.055 & \textbf{+0.018} & [+0.006, +0.030] \\
\midrule
Qwen3-8B & 3 & +0.161 & +0.028 & +0.005 & \textbf{+0.023} & [+0.015, +0.032] \\
 & 5 & +0.364 & +0.121 & +0.011 & \textbf{+0.110} & [+0.099, +0.122] \\
\midrule
R1-Distill-Qwen-14B & 3 & +0.416 & +0.073 & +0.007 & \textbf{+0.067} & [+0.053, +0.081] \\
 & 5 & +0.493 & +0.164 & +0.041 & \textbf{+0.123} & [+0.109, +0.140] \\
\midrule
Granite-3.3-8B & 3 & +0.379 & +0.067 & +0.077 & -0.010 & [-0.023, +0.003] \\
 & 5 & +0.267 & +0.089 & +0.092 & -0.003 & [-0.014, +0.007] \\
\midrule
GPT-OSS-20B & 3 & +0.266 & +0.047 & +0.005 & \textbf{+0.042} & [+0.032, +0.053] \\
 & 5 & +0.375 & +0.125 & +0.032 & \textbf{+0.093} & [+0.079, +0.108] \\
\midrule
   \bottomrule
 \end{tabular}
 \label{tab:matching}
\end{table*}

\subsection{The tags-only arm: roster tags without the decider's own group line}
\label{sm:halfarm}
The main study's visible arm shows the roster's tags and a line naming
the decider's own group together, so it cannot say which of them the
favoritism needs. The tags-only arm reported in Section~\ref{sec:halfarm} keeps the
roster tagged and withholds the line, changing nothing else in either prompt
(\texttt{src/probe.py --arm tagsonly}; the prompt builder in
\texttt{src/llm.py} documents the four-cell design this is part of). It was run
at minority sizes $3$ and $5$ for all six models at seeds $0$--$14$, which are
the first fifteen of the main study's thirty; a society's group memberships and
label pair are a pure function of the integer seed, so every comparison below
is paired by seed. Both label-swap passes were run, since a label still reaches
the prompt. That is $600$ trials per model per size, $7{,}200$ in all, plus a
partial run at the equal $10{:}10$ split ($460$ trials, three to four seeds per
model for three models) kept as the spot check that the rarity effect vanishes
when neither tag is rarer. The fourth cell of the design, the decider's own
line shown with the roster untagged, is implemented but was not run.

\emph{Favoritism} is still well defined in this arm: group membership is
assigned and recorded as usual, and the quantity is the contrast between
targets that do and do not share the decider's tag. What changes is that the
decider cannot identify that contrast. Table~\ref{tab:thirdarm} fits the main
text's dyad-level GEE in each arm at each collected size, every arm on all its
own seeds. Table~\ref{tab:rarity} is the seed-matched decomposition: pooled
over sizes $3$ and $5$ on the fifteen shared seeds, the visible arm's
minority$-$majority gap in excess share equals the tags-only arm's gap plus what the
own-group line adds, and the two parts are reported with a seed-level bootstrap
interval on the second ($10{,}000$ resamples).

At minority size $3$ the points a target receives by tag, rarer against common
where an even split gives $5.26$, are $13.0$ against $3.9$ (R1-Distill-Qwen-14B),
$7.5$ against $4.9$ (Granite-3.3-8B), $7.1$ against $4.9$ (GPT-OSS-20B), $7.1$
against $5.0$ (Qwen3-8B), $6.7$ against $5.0$ (R1-Distill-Llama-8B) and $5.3$
against $5.3$ (Phi-4-reasoning); at size $5$ the same ratios run $1.01$ to
$2.13$; at the equal split, Phi-4-reasoning and GPT-OSS-20B sit at
$5.2$--$5.3$ on either tag and R1-Distill-Qwen-14B at $5.5$ against $5.1$
(three to four seeds each).

Table~\ref{tab:halfarmcells} gives every cell of both arms on the shared
seeds, by decider group, with the own-group residual beside the ceiling $C$
that bounds an excess share. The residual is what Table~\ref{tab:rarity}
pools. It is positive in every cell and larger for minority deciders in
eleven of the twelve.

\paragraph{The decomposition is arithmetic, not mechanistic.} Visible minus tags-only is the own-group residual by definition, and
Section~\ref{sec:halfarm} reads it as what the line adds. That reading needs
the two arms to be additive: the rarity response the tags-only arm measures has to
be unchanged when the line is present. One cell shows that it is not.
R1-Distill-Qwen-14B's majority deciders at minority size~3 have a residual of
$+0.231$, above the $0.158$ that any majority decider can allocate in excess
and above the minority residual in the same societies ($+0.203$). No
allocation produces that. What produces it is the tags-only arm's rarity effect
($-0.222$) being far larger than anything the visible arm still shows
($+0.009$): in that model, being told its own tag does not add a component on
top of the rarity response but replaces most of it. The residual remains a
well-defined quantity, and Table~\ref{tab:rarity}'s claim that it is positive
in every model stands. What it is not, in general, is a measure of how hard a
decider allocates to its own group with the rarity effect held fixed.
Appendix~\ref{sm:normshare} divides an excess share by $C$; the residual is a
difference of two and must not be divided by it, which is how this cell was
first noticed, when an earlier version of the analysis produced a normalized
value of $1.107$.

\paragraph{The residual against the matching floor.}
Appendix~\ref{sm:matching} rules out size-independent matching for the visible
arm as measured. Section~\ref{sec:halfarm} then attributes part of the visible
arm's size dependence to the rarity effect, which the bound does not cover,
since it is not a match on the decider's tag. The question that leaves is
whether the residual is itself size-dependent, and under additivity it can be
asked in the same way: if the visible arm were a fixed-ratio match plus the
rarity effect, the residual would obey
$O_{\text{maj}} \ge C(n^{\text{maj}}_{\text{in}})/C(n^{\text{min}}_{\text{in}})\,O_{\text{min}}$.
Table~\ref{tab:residualmatching} evaluates it. The residual falls below the
floor in five of the twelve cells, and the interval excludes zero in two,
both Phi-4-reasoning's, the model in which the tags-only arm did least. In the
other five models no cell falls significantly below it, and in
R1-Distill-Qwen-14B and Granite-3.3-8B both cells sit well above it. The test
inherits the additivity assumption just discussed, so it is not decisive; but
it sets the limit on what Appendix~\ref{sm:matching} shows. The visible arm is
not the product of size-independent matching alone. Once the part the rarity
effect accounts for is removed, what remains is compatible with a fixed-ratio
match on the decider's tag in five of the six models, and
Section~\ref{sec:origin} claims no more than that.

\begin{table*}[t]
 \caption{Both arms, cell by cell, on the fifteen seeds they share.
 ``Visible'' and ``tags-only'' are the in-group excess share for deciders of
 that group at that minority size, the tags-only arm withholding the decider's own
 group line; ``residual'' is their difference, the own-group component of
 Table~\ref{tab:rarity} before pooling. $C=1-n_{\text{in}}/19$ is the largest
 excess share a decider of that cell can attain. The residual is a difference
 of two excess shares and is not bounded by $C$; the one cell in which it
 exceeds $C$ is marked ($\dagger$) and discussed in the text. Generated by
 \texttt{analysis/make\_rarity\_decomposition.py}.}
 \centering
 \small
 \begin{tabular}{l c l rr rrr}
   \toprule
   Model & Size & Decider & $n_{\text{in}}$ & $C$ & Visible & Tags-only & Residual \\
   \midrule
R1-Distill-Llama-8B & 3 & minority & 2 & 0.895 & +0.126 & +0.021 & +0.105 \\
 & 3 & majority & 16 & 0.158 & -0.037 & -0.046 & +0.009 \\
 & 5 & minority & 4 & 0.789 & +0.156 & +0.051 & +0.105 \\
 & 5 & majority & 14 & 0.263 & +0.011 & -0.043 & +0.054 \\
\midrule
Phi-4-reasoning (14B) & 3 & minority & 2 & 0.895 & +0.205 & +0.001 & +0.205 \\
 & 3 & majority & 16 & 0.158 & +0.019 & -0.002 & +0.021 \\
 & 5 & minority & 4 & 0.789 & +0.262 & +0.000 & +0.261 \\
 & 5 & majority & 14 & 0.263 & +0.056 & -0.002 & +0.059 \\
\midrule
Qwen3-8B & 3 & minority & 2 & 0.895 & +0.176 & +0.032 & +0.143 \\
 & 3 & majority & 16 & 0.158 & +0.004 & -0.054 & +0.058 \\
 & 5 & minority & 4 & 0.789 & +0.374 & +0.122 & +0.252 \\
 & 5 & majority & 14 & 0.263 & +0.012 & -0.060 & +0.072 \\
\midrule
R1-Distill-Qwen-14B & 3 & minority & 2 & 0.895 & +0.421 & +0.218 & +0.203 \\
 & 3 & majority & 16 & 0.158 & +0.009 & -0.222 & +0.231$^{\dagger}$ \\
 & 5 & minority & 4 & 0.789 & +0.499 & +0.194 & +0.304 \\
 & 5 & majority & 14 & 0.263 & +0.042 & -0.155 & +0.197 \\
\midrule
Granite-3.3-8B & 3 & minority & 2 & 0.895 & +0.369 & +0.052 & +0.317 \\
 & 3 & majority & 16 & 0.158 & +0.076 & -0.066 & +0.142 \\
 & 5 & minority & 4 & 0.789 & +0.292 & +0.081 & +0.212 \\
 & 5 & majority & 14 & 0.263 & +0.091 & -0.071 & +0.162 \\
\midrule
GPT-OSS-20B & 3 & minority & 2 & 0.895 & +0.229 & +0.044 & +0.186 \\
 & 3 & majority & 16 & 0.158 & -0.000 & -0.055 & +0.055 \\
 & 5 & minority & 4 & 0.789 & +0.380 & +0.061 & +0.318 \\
 & 5 & majority & 14 & 0.263 & +0.024 & -0.046 & +0.070 \\
\midrule
   \bottomrule
 \end{tabular}
 \label{tab:halfarmcells}
\end{table*}

\begin{table*}[t]
 \caption{The matching floor of Appendix~\ref{sm:matching} applied to the
 own-group residual. ``Residual'' is the visible-minus-tags-only excess share of
 Table~\ref{tab:halfarmcells}; ``floor'' is the least majority residual
 compatible with a size-independent match given the minority residual,
 $C(n^{\text{maj}}_{\text{in}})/C(n^{\text{min}}_{\text{in}})$ times it
 ($0.176$ at size~3, $0.333$ at size~5); ``shortfall'' is floor minus
 observed, with a $95\%$ percentile interval from resampling the shared seeds
 ($10{,}000$ resamples, both arms recomputed together). Bold: positive with an
 interval excluding zero, so the residual is below what any fixed ratio gives.
 A negative shortfall means the residual is at or above it. The test assumes
 the two arms are additive, which Table~\ref{tab:halfarmcells} shows fails in
 one cell.}
 \centering
 \small
 \begin{tabular}{l c rrr rl}
   \toprule
   Model & Size & Residual, minority & Floor & Residual, majority & Shortfall & $95\%$ CI \\
   \midrule
R1-Distill-Llama-8B & 3 & +0.105 & +0.018 & +0.009 & +0.009 & [-0.015, +0.035] \\
 & 5 & +0.105 & +0.035 & +0.054 & -0.019 & [-0.037, -0.002] \\
\midrule
Phi-4-reasoning (14B) & 3 & +0.205 & +0.036 & +0.021 & \textbf{+0.015} & [+0.005, +0.026] \\
 & 5 & +0.261 & +0.087 & +0.059 & \textbf{+0.029} & [+0.010, +0.047] \\
\midrule
Qwen3-8B & 3 & +0.143 & +0.025 & +0.058 & -0.033 & [-0.050, -0.014] \\
 & 5 & +0.252 & +0.084 & +0.072 & +0.012 & [-0.005, +0.028] \\
\midrule
R1-Distill-Qwen-14B & 3 & +0.203 & +0.036 & +0.231 & -0.195 & [-0.213, -0.179] \\
 & 5 & +0.304 & +0.101 & +0.197 & -0.095 & [-0.120, -0.069] \\
\midrule
Granite-3.3-8B & 3 & +0.317 & +0.056 & +0.142 & -0.086 & [-0.108, -0.068] \\
 & 5 & +0.212 & +0.071 & +0.162 & -0.091 & [-0.113, -0.068] \\
\midrule
GPT-OSS-20B & 3 & +0.186 & +0.033 & +0.055 & -0.022 & [-0.041, -0.003] \\
 & 5 & +0.318 & +0.106 & +0.070 & +0.036 & [-0.003, +0.072] \\
\midrule
   \bottomrule
 \end{tabular}
 \label{tab:residualmatching}
\end{table*}

\begin{table*}[t]
 \caption{The tags-only arm beside the arms it sits between. Dyad-level GEE
 (points $\sim$ same\_group, independence working correlation, seed-clustered
 robust SE) fit separately in each arm at each collected minority size, each on
 all of its own seeds ($30$ for visible and group-blind, $15$ for the tags-only arm).
 In the tags-only arm the decider is never told its own tag, so a negative
 coefficient is an over-allocation to the rarer tag read against the decider's
 unshown group: positive for minority deciders, negative for majority deciders,
 and negative pooled because majority deciders outnumber them. The group-blind
 floor is pooled over the same two sizes. Generated by
 \texttt{analysis/make\_thirdarm\_check.py}.}
 \centering
 \small
 \begin{tabular}{l c l rrrr r}
   \toprule
   Model & Size & Arm & $\beta$ & SE & $z$ & $p$ & dyads \\
   \midrule
R1-Distill-Llama-8B & 3 & visible & -0.379 & 0.130 & -2.9 & 0.0035 & 22,800 \\
 & 3 & tagsonly & -0.961 & 0.134 & -7.2 & $<10^{-12}$ & 11,400 \\
 & 5 & visible & +0.961 & 0.097 & +9.9 & $<10^{-22}$ & 22,800 \\
 & 5 & tagsonly & -0.428 & 0.132 & -3.2 & 0.0012 & 11,400 \\
 & 3+5 & hidden & +0.045 & 0.029 & +1.6 & 0.1181 & 22,800 \\
\midrule
Phi-4-reasoning (14B) & 3 & visible & +1.215 & 0.097 & +12.5 & $<10^{-35}$ & 22,800 \\
 & 3 & tagsonly & -0.049 & 0.021 & -2.3 & 0.0208 & 11,400 \\
 & 5 & visible & +2.123 & 0.108 & +19.6 & $<10^{-84}$ & 22,800 \\
 & 5 & tagsonly & -0.038 & 0.010 & -3.7 & 0.0003 & 11,400 \\
 & 3+5 & hidden & +0.006 & 0.005 & +1.1 & 0.2665 & 22,800 \\
\midrule
Qwen3-8B & 3 & visible & +0.763 & 0.110 & +6.9 & $<10^{-11}$ & 22,800 \\
 & 3 & tagsonly & -1.106 & 0.112 & -9.9 & $<10^{-22}$ & 11,400 \\
 & 5 & visible & +2.181 & 0.114 & +19.1 & $<10^{-80}$ & 22,800 \\
 & 5 & tagsonly & -0.313 & 0.108 & -2.9 & 0.0036 & 11,400 \\
 & 3+5 & hidden & +0.009 & 0.008 & +1.2 & 0.2288 & 22,800 \\
\midrule
R1-Distill-Qwen-14B & 3 & visible & +1.823 & 0.176 & +10.4 & $<10^{-24}$ & 22,800 \\
 & 3 & tagsonly & -4.183 & 0.197 & -21.2 & $<10^{-99}$ & 11,400 \\
 & 5 & visible & +3.389 & 0.124 & +27.4 & $<10^{-164}$ & 22,800 \\
 & 5 & tagsonly & -1.490 & 0.238 & -6.3 & $<10^{-9}$ & 11,400 \\
 & 3+5 & hidden & -0.042 & 0.045 & -0.9 & 0.3470 & 22,800 \\
\midrule
Granite-3.3-8B & 3 & visible & +3.278 & 0.087 & +37.8 & $<10^{-300}$ & 22,800 \\
 & 3 & tagsonly & -1.296 & 0.180 & -7.2 & $<10^{-12}$ & 11,400 \\
 & 5 & visible & +2.994 & 0.110 & +27.3 & $<10^{-163}$ & 22,800 \\
 & 5 & tagsonly & -0.724 & 0.121 & -6.0 & $<10^{-8}$ & 11,400 \\
 & 3+5 & hidden & +0.050 & 0.032 & +1.6 & 0.1172 & 22,800 \\
\midrule
GPT-OSS-20B & 3 & visible & +1.185 & 0.117 & +10.1 & $<10^{-23}$ & 22,800 \\
 & 3 & tagsonly & -1.080 & 0.169 & -6.4 & $<10^{-9}$ & 11,400 \\
 & 5 & visible & +2.586 & 0.097 & +26.8 & $<10^{-157}$ & 22,800 \\
 & 5 & tagsonly & -0.428 & 0.181 & -2.4 & 0.0182 & 11,400 \\
 & 3+5 & hidden & -0.015 & 0.051 & -0.3 & 0.7659 & 22,800 \\
\midrule
   \bottomrule
 \end{tabular}
 \label{tab:thirdarm}
\end{table*}

\begin{table*}[t]
 \caption{The size asymmetry split into its two sources. ``Visible gap'' is the
 visible arm's minority$-$majority difference in in-group excess share, pooled
 over minority sizes $3$ and $5$ on the $15$ seeds shared with the tags-only arm;
 ``rarity gap'' is the same quantity in the tags-only arm, where the decider's own
 group line is withheld; ``own-group gap'' is their difference, what the line
 adds, with a $95\%$ seed-level bootstrap interval ($10{,}000$ resamples of the
 shared seeds, both arms recomputed together) and shown bold where the interval
 excludes zero. ``Rarity share'' is the rarity gap as a fraction of the visible
 gap. Fig.~\ref{fig:rarity} plots the same values. Generated by
 \texttt{analysis/make\_rarity\_decomposition.py}.}
 \centering
 \small
 \begin{tabular}{l c rr r l r}
   \toprule
   Model & Seeds & Visible gap & Rarity gap & Own-group gap & $95\%$ CI & Rarity share \\
   \midrule
R1-Distill-Llama-8B & 15 & +0.153 & +0.080 & \textbf{+0.073} & [+0.035, +0.113] & 52\% \\
Phi-4-reasoning (14B) & 15 & +0.196 & +0.003 & \textbf{+0.193} & [+0.164, +0.222] & 1\% \\
Qwen3-8B & 15 & +0.267 & +0.134 & \textbf{+0.133} & [+0.083, +0.189] & 50\% \\
R1-Distill-Qwen-14B & 15 & +0.435 & +0.395 & \textbf{+0.040} & [+0.002, +0.074] & 91\% \\
Granite-3.3-8B & 15 & +0.247 & +0.135 & \textbf{+0.112} & [+0.069, +0.154] & 55\% \\
GPT-OSS-20B & 15 & +0.293 & +0.103 & \textbf{+0.189} & [+0.129, +0.249] & 35\% \\
   \bottomrule
 \end{tabular}
 \label{tab:rarity}
\end{table*}

\subsection{The R1-Distill-Llama-8B majority-decider exception at minority size 3}
\label{sm:reversal}
Appendix~\ref{sm:stats} and Table~\ref{tab:share} report one cell that departs from
the otherwise uniform pattern of the main text: for R1-Distill-Llama-8B, majority
deciders at the smallest minority size ($3$ of $20$) show significantly
\emph{negative} favoritism (points per target $-1.45$; excess share $-0.037$; both
$p<10^{-8}$), i.e.\ these majority deciders give the numerically small out-group
\emph{more} than a per-capita-even split would predict. Two checks establish that
this is a real, if narrow, effect and not an artifact of parsing or aggregation.

First, at the trial level the raw allocations behind this cell are well-formed,
correctly summed JSON objects, not parse failures: a subset of majority deciders
give nearly the entire $100$-point pool to the three out-group targets specifically
(e.g.\ one trial assigns $34$, $33$ and $33$ points to the three minority targets
and $0$ to all sixteen majority targets), while the remaining majority deciders in
this cell split points close to evenly. Second, and more decisively, the effect is
not confined to a handful of extreme trials: aggregating to the seed level (the
paper's unit of analysis, averaging over a society's majority deciders and both
label-swap passes), \textbf{every one of the $30$ independent societies} shows
negative mean favoritism for majority deciders in this cell ($30/30$; one-sided
sign test $p<10^{-9}$) -- this is a population-level, not a tail-driven, reversal.

Table~\ref{tab:seedsign} disaggregates this seed-level sign count by minority size
for all six models, which Table~\ref{tab:prevalence}'s pooled-across-size
prevalence obscures. The $30/30$ reversal is unique to R1-Distill-Llama-8B at
minority size $3$: it is absent at size $5$ ($12/30$, not significant) and size
$10$ ($0/30$), absent in every other model at every size (at most $11/30$,
GPT-OSS-20B at size $3$; none significant), and the group-blind floor for this same
cell sits at $10/30$, a count the one-sided binomial does not separate from chance.
Pooling the group-blind arm over sizes and decider groups puts the statistic's null
behavior on a broader base: its positive counts there run $34/90$ to $55/90$
across the six models, around the $45/90$ expected under no bias, so the sign count
is well behaved and the $30/30$ is not an artifact of it. The tags-only arm settles what the cell is (Appendix~\ref{sm:halfarm}). With the
decider's own group line withheld, this model's majority deciders at minority
size $3$ still give the three rare-tagged targets more than their share, at
$-0.046$ excess share; in the visible arm the same cell reads $-0.037$, so the
line adds only $+0.009$. The reversal is the rarity effect of
Section~\ref{sec:halfarm} in the one cell where it outweighs the own-group
component -- the same two components every other cell shows, with the balance
tipped -- and not a behavior specific to this model. The paraphrase check bounds the cell. Under the reworded instrument this
cell's favoritism is $+0.25$ points per target rather than $-1.45$, and the
same model's group-blind floor for the matching condition is $+0.30$
(Table~\ref{tab:paraphrase}), so read against the instrument that produced it,
as every value in this paper is, the cell sits at its floor ($-0.05$). The
reversal is therefore a property of this model under the original wording and
not a wording-independent feature: the balance between the rarity and
own-group components that produces it is itself wording-sensitive, as the
visible effects are (Appendix~\ref{sm:paraphrase}).
We read the exception, then, as real under the original wording, narrow and
accounted for: the rarity
effect is largest at the most lopsided split and this model's own-group
component among majority deciders at that split is the smallest of the six, so this is the one cell where the first
outweighs the second. It is not a failure of the instrument, and it tempers how
the main text's ``close to proportionally'' characterization should be read for
majority deciders generally, since Section~\ref{sec:halfarm} shows that
near-proportional net allocation is own-group favoring offset by the rarity
effect rather than its absence. It is not exclusive to the R1-distillation
lineage: smaller, non-significant counts in the same direction appear in
R1-Distill-Qwen-14B ($10/30$) and, slightly larger still, in GPT-OSS-20B ($11/30$),
both at the same size and both far from the $30/30$ seen in R1-Distill-Llama-8B.

\begin{table*}[t]
 \caption{Seed-level sign count of \emph{negative} mean favoritism among majority
 deciders, disaggregated by minority size (visible arm; Benjamini--Hochberg
 correction not applied since this is a single-cell follow-up, not part of the
 72-test family in Table~\ref{tab:fdr}). ``$k/30$ ($p$)'' is the number of the $30$
 societies with negative mean favoritism for majority deciders at that size, with a
 one-sided binomial test against $0.5$; a large, significant count indicates a
 population-level reversal (out-group favored over in-group) rather than the
 modal near-zero-or-positive pattern. The last column is the same count at the
 group-blind floor for minority size $3$, for comparison.}
 \centering
 \small
 \begin{tabular}{l ccc c}
   \toprule
   & \multicolumn{3}{c}{Labels visible, minority size} & Group-blind (size 3) \\
   \cmidrule(lr){2-4}
   Model & $3$ & $5$ & $10$ & \\
   \midrule
R1-Distill-Llama-8B & 30/30 ($<10^{-9}$) & 12/30 (0.8998) & 0/30 (1.0000) & 10/30 \\
Phi-4-reasoning (14B) & 2/30 (1.0000) & 0/30 (1.0000) & 0/30 (1.0000) & 15/30 \\
Qwen3-8B & 9/30 (0.9919) & 7/30 (0.9993) & 0/30 (1.0000) & 13/30 \\
R1-Distill-Qwen-14B & 10/30 (0.9786) & 4/30 (1.0000) & 0/30 (1.0000) & 14/30 \\
Granite-3.3-8B & 0/30 (1.0000) & 0/30 (1.0000) & 0/30 (1.0000) & 7/30 \\
GPT-OSS-20B & 11/30 (0.9506) & 1/30 (1.0000) & 0/30 (1.0000) & 12/30 \\
   \bottomrule
 \end{tabular}
 \label{tab:seedsign}
\end{table*}

\section{Matched reasoning-mode control}
\label{sm:reasoning}
The six-model sample in the main text is reasoning-only, so it cannot on its own
attribute the effect to deliberation versus any other property of a model
family. We therefore reran the identical probe on Qwen3-8B with its
\texttt{thinking} mode disabled -- the same checkpoint, weights, and training
pipeline as the Qwen3-8B row of Table~\ref{tab:gee}, differing only in whether the
model deliberates in an extended chain of thought before answering
(\texttt{enable\_thinking=False}; Appendix~\ref{sm:models}). This is a within-checkpoint control, not
a second independent model: it isolates the reasoning/non-reasoning contrast from
the architecture, pretraining corpus, and post-training recipe that differ across
the other five models in the main sample.

The design and scale match the main study exactly: the full $3\times2$
minority-size$\times$visibility design, 30 seeds per cell, 5{,}400 trials. Parse
success ranged 98.3--100\% across the six conditions (Table~\ref{tab:parse}'s range
for the six reasoning models is 98.6--99.8\%), and the group-blind floor is clean in
both modes (thinking-enabled $0.011\pm0.008$; non-thinking $-0.003\pm0.004$
points per target; both $p>0.13$), so the instrument validates equally without
reasoning.

Table~\ref{tab:reasoning} reports the dyad-level GEE comparison. With labels
visible, non-thinking Qwen3-8B shows \emph{more} overall in-group favoritism than
its thinking-enabled counterpart ($3.81$ vs.\ $2.31$ points per target;
same-group$\times$reasoning interaction $-1.50$, $p=3\times10^{-83}$) -- disabling
reasoning did not remove the effect here, if anything the opposite. But the
minority$-$majority gap collapses from $+4.08$ points per target (matching
Table~\ref{tab:gee}) to $+0.54$ ($p=0.002$; an order of magnitude smaller, though
still distinguishable from zero; same-group$\times$reasoning$\times$decider-group
interaction $+3.55$, $p=3\times10^{-63}$). The size-robust excess-share measure
(pooled over minority sizes $3$ and $5$) shows the same asymmetry collapse:
thinking-enabled majority deciders allocate close to proportionally ($+0.01$ excess
share, matching the main text's Qwen3-8B row), while non-thinking majority deciders
allocate nearly as much excess share to their own group as minority deciders do
($+0.15$ vs.\ $+0.17$, compared with $+0.01$ vs.\ $+0.26$ under thinking). In other
words, disabling reasoning does not make majority deciders less own-group-favoring in
absolute terms -- if anything favoritism is more pervasive overall -- but it removes
the distinction between how minority and majority deciders behave, which is what
drives the asymmetry to near zero.

\begin{figure*}[t]
  \centering
  \includegraphics[width=\textwidth]{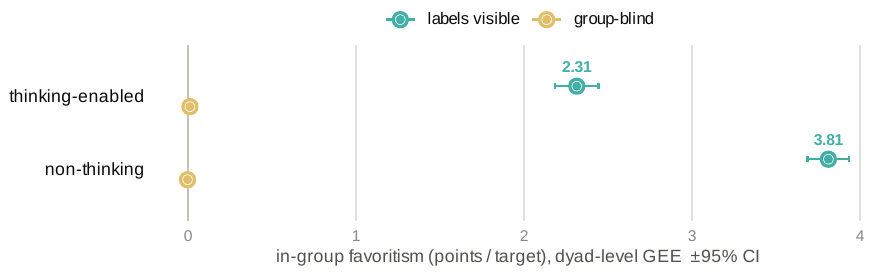}
  \caption{\textbf{Reasoning and the minority asymmetry: a matched Qwen3-8B
  control.} Dyad-level GEE estimates (points per target, $\pm$95\% CI, seed-clustered
  robust SE) for the same Qwen3-8B checkpoint run with thinking enabled versus
  disabled, with labels visible (teal) and in the group-blind control (gold). Both
  modes show a favoritism effect far from the floor; the modes' relative sizes and
  the minority$-$majority breakdown behind them are reported in
  Table~\ref{tab:reasoning}.}
  \label{fig:reasoning}
\Description{A forest plot of dyad-level estimates of in-group favoritism in points per target with 95 percent confidence intervals for one Qwen3-8B checkpoint run with thinking enabled and with thinking disabled, under labels-visible and group-blind conditions. Both visible estimates sit far above zero, the non-thinking one higher than the thinking-enabled one, while both group-blind estimates sit on zero.}
\end{figure*}

Figure~\ref{fig:reasoning} plots these dyad-level estimates with their 95\%
intervals, scoped to this one model's two reasoning modes.

\begin{table*}[t]
 \caption{Matched reasoning-mode control: Qwen3-8B, thinking-enabled versus
 non-thinking (same checkpoint, weights, and training pipeline; dyad-level GEE,
 points per target, seed-clustered robust SE; $z=\beta/\text{SE}$). The
 same-group$\times$reasoning interaction is the shift in overall favoritism from
 disabling reasoning; the triple interaction with decider-group is the shift in the
 minority$-$majority asymmetry specifically. Rows 1 and 4 reproduce the
 same\_group coefficients reported for the thinking-enabled arm in
 Table~\ref{tab:gee}; rows 2 and 5 are the corresponding non-thinking fits.}
 \centering
 \small
 \begin{tabular}{l rrrr r}
   \toprule
   Comparison & $\beta$ & SE & $z$ & $p$ & dyads \\
   \midrule
in-group favoritism, thinking-enabled (visible) & +2.314 & 0.066 & +35.1 & $<10^{-269}$ & 68,400 \\
in-group favoritism, non-thinking (visible) & +3.812 & 0.063 & +60.4 & $<10^{-300}$ & 68,400 \\
same-group $\times$ reasoning interaction (visible) & -1.498 & 0.077 & -19.3 & $<10^{-82}$ & 136,800 \\
in-group favoritism, thinking-enabled (group-blind) & +0.011 & 0.008 & +1.5 & 0.1367 & 34,200 \\
in-group favoritism, non-thinking (group-blind) & -0.003 & 0.004 & -0.8 & 0.4261 & 34,200 \\
same-group $\times$ reasoning interaction (group-blind) & +0.014 & 0.008 & +1.8 & 0.0741 & 68,400 \\
minority$-$majority gap, thinking-enabled (visible) & +4.083 & 0.183 & +22.3 & $<10^{-109}$ & 68,400 \\
minority$-$majority gap, non-thinking (visible) & +0.537 & 0.173 & +3.1 & 0.0019 & 68,400 \\
same-group $\times$ reasoning $\times$ decider-group (visible) & +3.546 & 0.211 & +16.8 & $<10^{-62}$ & 136,800 \\
   \bottomrule
 \end{tabular}
 \label{tab:reasoning}
\end{table*}

This is a single matched pair from a single model family, and we read it
accordingly: it rules out the simplest story (that reasoning is necessary for the
effect to appear, or that removing it uniformly shrinks favoritism), but it
does not establish that reasoning mode has this same effect on the
minority-majority asymmetry in the other five models, nor that the
pattern generalizes to non-reasoning models as a class. A within-study
thinking-enabled-versus-disabled contrast across every family that exposes the
switch, together with models never trained to deliberate at all, is the natural
next step (Section~\ref{sec:limitations}).

\end{document}